\documentclass[preprint]{aastex}
\usepackage{amsmath}
\usepackage{xspace}

\def\ie{i.e.\xspace}

\def\ltsima{$\; \buildrel < \over \sim \;$}
\def\simlt{\lower.5ex\hbox{\ltsima}}
\def\gtsima{$\; \buildrel > \over \sim \;$}
\def\simgt{\lower.5ex\hbox{\gtsima}}
\def\fesc{{$\langle f_{esc}\rangle$}\xspace}

\def\H2{H$_2$\xspace} 
\def\m{$^{-1}$\xspace}
\def\mm{$^{-2}$\xspace}
\def\mmm{$^{-3}$\xspace}
\def\pp{$^2$\xspace}

\def\ion#1#2{\text{#1\,\sc #2}}
\def\HI{{\ion{H}{i} }}
\def\HII{{\ion{H}{ii} }}
\def\GI{{\ion{He}{i} }}
\def\GII{{\ion{He}{ii} }}

\def\popII{Population~II\xspace}
\def\pop3{Population~III\xspace}
\def\pII{``large-halo''\xspace}
\def\p3{``small-halo''\xspace}
\def\pp3{``Small-halo''\xspace}
\def\Mpc{$h^{-1}$~Mpc\xspace}

\def\pc{$h^{-1}$~pc\xspace}
\def\Ms{$h^{-1}$~M$_\odot$\xspace}
 
\received{.......}
\accepted{.......}
\slugcomment{to be submitted to ApJ }

\begin{document}
\tighten
\thispagestyle{empty}

\pagestyle{myheadings}
\markright{DRAFT: \today\hfill}

\def\placefig#1{#1}

\title{THE FATE OF THE FIRST GALAXIES. I.\\
SELF-CONSISTENT COSMOLOGICAL SIMULATIONS WITH RADIATIVE TRANSFER} 

\author{MASSIMO RICOTTI, NICKOLAY Y. GNEDIN AND J. MICHAEL SHULL$^1$ } 

\affil{
Center for Astrophysics and Space Astronomy \\
Department of Astrophysical and Planetary Sciences \\
University of Colorado, Campus Box 389, Boulder CO 80309 \\
E--mail: ricotti, gnedin, mshull@casa.colorado.edu \\ 
$^1$ also at JILA, University of Colorado and National
Institute of Standards and Technology }  
 
\begin{abstract}
  In cold dark matter (CDM) cosmogonies, low-mass objects play an
  important role in the evolution of the universe. Not only are they
  the first luminous objects to shed light in a previously dark
  universe, but, if their formation is not inhibited by their own
  feedback, they dominate the galaxy mass function until redshift $z
  \sim 5$. In this paper we present and discuss the implementation of
  a 3D cosmological code that includes most of the needed physics to
  simulate the formation and evolution of the first galaxies with a
  self-consistent treatment of radiative feedback. The simulation
  includes continuum radiative transfer using the ``Optically Thin
  Variable Eddington Tensor'' (OTVET) approximation and line-radiative
  transfer in the \H2 Lyman-Werner bands of the background UV radiation.
  We include detailed chemistry for H$_2$ formation/destruction,
  molecular and atomic cooling/heating processes, ionization by
  secondary electrons, and heating by Ly$\alpha$ resonant scattering.
  
  We find that the first galaxies (``small-halos'') are characterized
  by bursting star formation, self-regulated by a feedback process
  that acts on cosmological scales. The mass in stars produced by
  these objects can exceed the mass in stars produced by normal
  galaxies; therefore, their impact on cosmic evolution cannot be
  neglected. The main focus of this paper is on the methodology of the
  simulations, and we only briefly introduce some of the results.  An
  extensive discussion of the results and the nature of the feedback
  mechanism are the focus of a companion paper.
 
\end{abstract}

\section{Introduction \label{sec:int}}

In the last five years, precise observations of the cosmic microwave
background (CMB), Ly$\alpha$ forest, galaxy clusters, and large
scale structure have produced a large amount of data that strongly
constrain the standard cosmological model. Until now, despite some
unsolved problems \citep*{Moore:94, Klypin:99, Navarro:00}, cold dark
matter (CDM) cosmogonies remain the most predictive cosmological models.
In CDM cosmogonies, the large galaxies that we observe today are the
result of the mergers of a large number of small and faint galaxies
(dwarf galaxies). The number of dwarf galaxies observed in the local
universe is larger than the number of normal galaxies but, today, they
account for only a small fraction of the mass in collapsed objects.
About 12 billion years ago, galaxies with the mass of our Milky Way
were extremely rare, and the universe was filled with objects similar
to dwarf spheroidal galaxies.

How small can a galaxy be? Or how small can a protogalaxy be in order
to be able to form stars? A necessary condition for star formation is
that the gas in the protogalaxy must cool to a temperature lower than
the virial temperature of the halo in a Hubble time. If the dark
matter (DM) mass is $M_{DM} \simgt 10^8$ M$_\odot$, the gas in the
protogalaxy (a so-called \pII protogalaxy) cools by \HI emission lines
because the virial temperature of the halo is $T_{vir} \simgt 10^4$ K.
But if $M_{DM} \simlt 10^8$ M$_\odot$ and the protogalaxy is of
primordial composition, the gas cannot cool ($T_{vir} \simlt 10^4$ K)
unless molecular hydrogen (\H2) is present. We call such an object a
\p3 protogalaxy. During the virialization process, the gas is
partially ionized ($x_e \sim 10^{-4}$) and heated to the virial
temperature by shocks. Free electrons can be captured by neutral
hydrogen (\HI) to produce H$^{-}$, which is the main catalyst for \H2
formation in a gas of primordial composition \citep{Lepp:84}. The
chemical reaction \HI$+{\rm H}^{-} \rightarrow {\rm H}_2 + e^-$
produces enough \H2 ($x_{H_2} \sim 10^{-4}$) to allow the formation of
galaxies as small as $M_{DM} \sim 10^5$ M$_\odot$ at $z \sim 30$, \ie,
with virial temperature $T_{vir} \sim 360$ K \citep*{Tegmark:97,
  Abel:00}.

\pp3 objects are thought to be the first galaxies formed in the
universe and, if their formation is not inhibited by feedback, they
should account for the bulk of mass in stars until reionization at $z
\sim 6$.  In Figure~\ref{fig:ps} we show the fraction of collapsed
(virialized) DM as a function of the DM halo mass at $z=24, 19, 16, 9,
5, 3, 0$ calculated with the Press-Schechter formula
\citep*{Press:74}. The inserted table shows the fraction of the
collapsed mass in \p3 objects ($T_{vir}<10^4$ K), normal galaxies
(\pII objects with $10^4~{\rm K}\le T_{vir} \le 10^6$ K), and clusters
($T_{vir}>10^6$ K).  The last column in the table shows the ratio of
the collapsed mass of the \p3 to the \pII objects. The hexagons,
squares, and triangles show the mass of 1 $\sigma$, 2 $\sigma$, and 3
$\sigma$ perturbations of the initial density field, respectively.
\pp3 objects are the dominant fraction of the DM collapsed mass before
redshift $z \sim 9$, and they remain an important fraction of the DM
collapsed mass until redshift $z \sim 5$.

\def\capfiga{%
  Fraction of collapsed (virialized) DM as a function of
  the DM halo mass at $z=25$, calculated using the Press-Schechter
  formalism. The inserted table shows the fraction of the collapsed
  mass in \p3 protogalaxies ($T_{vir}< 10^4$ K), normal galaxies (\pII
  protogalaxies with $10^4 {\rm K} \le T_{vir} \le 10^6$ K), and
  clusters ($T_{vir}>10^6$ K). The last column in the table shows
  the ratio of the collapsed mass of \p3 to \pII protogalaxies.
  \pp3 objects are an important fraction of the DM collapsed mass until
  redshift $z \sim 5$. Each curve has two thick portions: the thick
  section on the left shows \p3 objects, and the thick section on the right,
  clusters. In between, according to our definition, are
  \pII objects.}
\placefig{
\begin{figure*}[thp]
\plotone{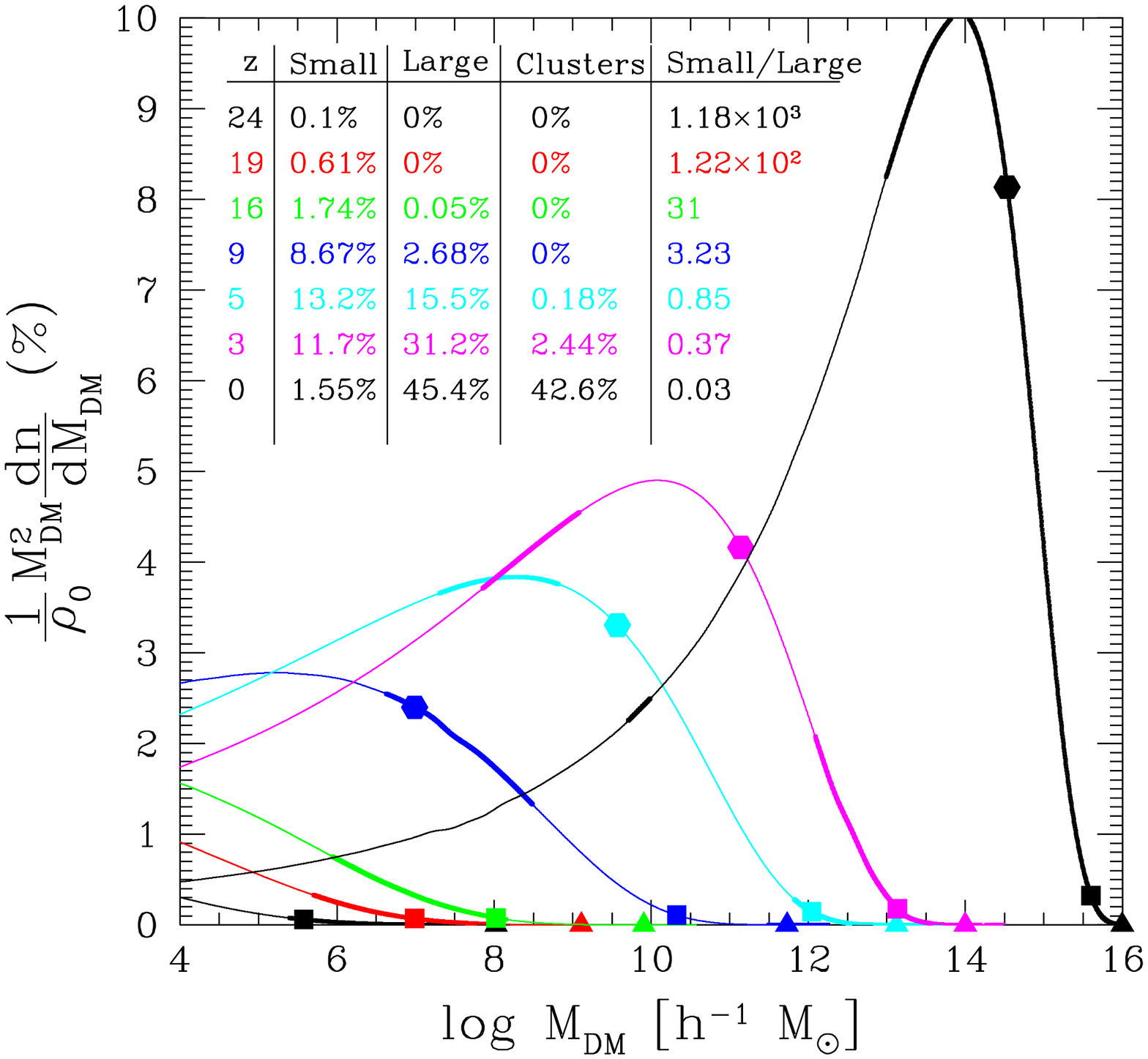}
\caption{\label{fig:ps}\capfiga}
\end{figure*}
}

Competing feedback effects determine the fate of \p3 objects.
Radiative feedback regulates the formation and destruction of \H2 in
the protogalaxies. Mechanical and thermal feedback from supernova (SN)
explosions could blow away the interstellar medium of \p3 objects,
produce \H2 \citep*{Ferrara:98}, or destroy \H2. In this paper, we
focus on radiative feedback processes. From the first works on this
subject \citep*{HaimanRL:97, HaimanAR:00, CiardiF:00, Machacek:00},
the negative feedback from the background in the \H2 Lyman-Werner
bands was thought to be the main process that regulates the formation
of \p3 objects.  The background in the Lyman-Werner bands dissociates
\H2 through the two-step Solomon process, suppressing or delaying the
formation of \p3 objects.  But there are also positive feedback
processes that can enhance the formation rate of \H2
\citep*{MacLow:86, ShapiroKang:87, HaimanRL:96, Ferrara:98, Oh:00,
  RicottiGS:01}. It is not trivial to determine which feedback
prevails by means of semi-analytic models.  Are \p3 objects able to
form and survive the negative feedback effect of the dissociating
background, or is their number drastically suppressed?  Are \p3
objects numerous enough to have some effect on the subsequent
evolution of the IGM? In this work we try to answer these questions,
using self-consistent 3D cosmological simulations with radiative
transfer.

We present our results in two papers.  This first paper focuses on the
implementation of the code and its convergence. In
\cite*{RicottiGSb:02} (Paper~II) we show and discuss the results.

The simulations we present here are the first 3D cosmological
simulations that include a self-consistent treatment of galaxy
formation with radiative transfer by following the evolution of DM
particles, star particles, gas, and photons. We also solve the line
radiative transfer for the background in the Lyman-Werner bands. A
number of potentially relevant physical processes are included as
well: secondary ionizations of H and He, detailed \H2 chemistry and
cooling processes, heating by Ly$\alpha$ resonant scattering, H and He
recombination lines, heavy element production and radiative cooling,
and the spectral energy distribution (SED), $g_\nu$, of the stellar
sources consistent with the choice of the Lyman-continuum (Lyc) escape
fraction, \fesc, from the resolution element of the simulation. These
ingredients, together with careful convergence studies, allow us to
study the formation and feedback of \p3 objects by means of numerical
simulations.  The advantage of this approach, compared to
semi-analytic models, is that we make no assumptions about the
physical processes that are important in the simulation, apart from
sub-grid physics. We assume that, when the gas sinks below the
resolution limit of the simulation, it forms stars according to the
Schmidt law (see eq.~[\ref{eq:sf}]).

The paper is organized in the following manner. In \S~\ref{sec:meth}
we describe the cosmological code and the detailed physical processes
simulated.  In \S~\ref{sec:resol} we study the convergence of the
simulations, and in \S~\ref{sec:sumI} we summarize the features of the
code and its limitations. We also provide a few examples of the
results to demonstrate the power of the code to for addressing
astrophysical problems.

\section{The Code \label{sec:meth}}

The simulations presented in this paper were performed with the
``Softened Lagrangian Hydrodynamics'' (SLH-P$^3$M) code described in
detail in \cite*{Gnedin:95, Gnedin:96a}, \cite*{GnedinB:96,
  GnedinOstraiker:97}, and \cite*{GnedinA:01}. The code solves the
system of time-dependent equations of motion of four main components:
dark matter particles (P$^3$M algorithm); gas particles
(quasi-Lagrangian deformable mesh using the SLH algorithm);
``star-particles'' formed using the Schmidt law in resolution elements
that sink below the numerical resolution of the code; and photons
solved self-consistently with the radiative transfer equation in the
OTVET approximation \citep{GnedinA:01}. We solve the line radiative
transfer of the background radiation in the \H2 Lyman-Werner bands
with spectral resolution $\Delta \nu /\nu= 9.26 \times 10^{-6}$ (\ie,
20,000 logarithmic bins in the energy range $11.2-13.6$ eV of the \H2
Lyman-Werner bands). Detailed atomic and molecular physics is
included: secondary ionizations of H and He, \H2 chemistry and cooling
processes, heating by Ly$\alpha$ resonant scattering, H and He
recombination lines, heavy element production and radiative cooling,
and galaxy spectrum consistent with the value of \fesc.

For clarity, we first summarize some relevant features of the code
already implemented and published. The new components of the code are
described in more detail as separate subsections.

\subsubsection{Radiative transfer: the OTVET approximation}

The OTVET approximation is based on solving equations for the first
two moments of the photon distribution function.  Since the moment
equations have the conservative form, the photon number and flux
conservation are achieved automatically in this approximation.
However, the hierarchy of moment equations is not closed at any finite
level, i.e.\ the flux (first moment) equation contains the second
moment (flux tensor), which can be reduced to the Variable Eddington
tensor (the latter has a unit trace).  In the OTVET approximation, the
Eddington tensor is calculated in the optically thin approximation -
hence the abbreviation OTVET: Optically Thin Variable Eddington Tensor
approximation.  Since the two moment equations can be solved very fast
numerically, most of the computational time is spent in calculating
the Eddington tensor, which can be done using the same algorithm used
for computing gravity, in our case the Adaptive P$^3$M.

Methodically, this approach is equivalent to separating the algorithm
for calculating absorption and emission of photons from the algorithm
for calculating the direction of propagation of a photon. The first
part can be done rapidly and accurately by using moments of the
distribution function.  The second part, in a direct ray-tracing
approach, will be unacceptably slow.  The OTVET approximation allows
us to speed up this calculation by an enormous factor without
violating the accuracy of the first algorithm.

Another feature of our implementation of the OTVET approximation is
the use of ``effective'' frequencies. Accurate computation of
photoionization and photoheating rates on a logarithmically spaced
mesh in the frequency space requires about 20 points per each
e-folding, or about 300 frequency bins altogether. Computing the
radiative transfer for each of these bins is impractical at the
moment. However, since we are ultimately only interested in the
integrals over the frequencies, we can speed up the calculation of the
reaction rates by a factor of about 50 by adopting the following
ansatz for the radiation energy density $I_\nu$ as a function of
frequency:
\[
I_\nu = I_{\nu,\rm OT} \exp\left(-\sum_\alpha \sigma_\nu^{(\alpha)}      
N_{\rm eff}^{(\alpha)} \right),
\]         
where index $\alpha$ runs over the list of species, which includes
\HI, \GI, and \GII (since we treat H$_2$ in the optically thin
approximation), $I_{\nu,\rm OT}$ is the radiation energy density
calculated in the optically thin regime, and the column densities,
$N_{\rm eff}^{(\alpha)}$, are functions of position and time only.
Thus, in order to compute $N_{\rm eff}^{(\alpha)}$, we need to solve
the radiative transfer equation in the optically thin regime (which is
equivalent to just one frequency since the frequency dependence
factors out) and at three other frequencies which we choose to be just
above the \HI, \GI, and \GII ionization thresholds.

\subsubsection{Heavy elements}

In this version of the code we do not include the mechanical energy
input from SN explosions, but we do account for the metal enrichment
produced by SNe. The cooling function is calculated according to the
gas metallicity.  Since it is impractical to treat detailed ionization
and thermal balance for some 130 ionization states of most common
heavy elements (metals), we assume that whenever metal cooling is
important, the different ionization states of heavy elements are in
ionization equilibrium. In that case, we can adopt a single cooling
function that describes the cooling rate due to excitation of atomic
transitions in heavy elements. The metallicity of the stars produced
in the simulation are recorded, but the emission spectrum is not
calculated consistently with the metallicity of the stellar
population.

\subsubsection{Cosmological model}

We adopt a $\Lambda$CDM cosmological model with the following
parameters: $\Omega_0 = 0.3$, $\Omega_\Lambda= 0.7$, $h=0.7$, and
$\Omega_b = 0.04$. The initial conditions at $z = 100$, where we start
our simulations, are computed using the COSMICS package
\citep*{Bertschinger:95} assuming a spectrum of dark matter
perturbations with power-law slope $n=1$ and COBE normalization.  We
parameterize the unknown sub-grid physics of star formation at high
redshift with three free parameters: the star formation efficiency,
$\epsilon_*$, the energy fraction in emitted ionizing photons per
baryon converted into stars, $\epsilon_{UV}$, and the Lyman continuum
(Lyc) escape fraction, \fesc.

In the rest of this section, we will describe in detail some new
components included in the code that are particularly useful to
address the problem of the formation of the first luminous objects.

\subsection{Line Radiative Transfer in the Lyman-Werner Bands\label{ssec:LW}}

The evolution of the specific intensity $J_\nu$ [erg cm\mm s\m Hz\m sr\m]
of ionizing or dissociating radiation in the expanding universe, with no scattering, is given by the following equation:
\begin{equation}
        {\partial J_\nu\over\partial t} +
        {\partial\over\partial x^i}\left(\dot{x}^i J_\nu\right) -
        H\left(\nu{\partial J_\nu\over\partial\nu}-3J_\nu\right) =
        - k_\nu J_\nu + S_\nu.
        \label{eq:Jnueq}
\end{equation}
Here, $x^i$ are the comoving coordinates, $H$ is the Hubble constant,
$k_\nu$ is the absorption coefficient, $S_\nu$ is the source function,
and $\dot{x}^i=cn^i/a$, where $n^i$ is the unit vector in the
direction of photon propagation and $a=(1+z)^{-1}$ is the scale factor.
The volume-averaged mean specific intensity (``background'') is
\begin{equation}
        \bar J_\nu(t) \equiv \langle J_\nu(t,\vec{x},\vec{n})\rangle_V,
        \label{eq:Jnubardef}
\end{equation}
where the averaging operator acting on a function $f(\vec{x},\vec{n})$
of position and direction is defined as:
\begin{equation}
        \langle f(\vec{x},\vec{n})\rangle_V =
        \lim_{V\rightarrow\infty} {1\over 4\pi V}
        \int_V d^3x \int d\Omega f(\vec{x},\vec{n}).
        \label{averagoper}
\end{equation}
The mean intensity $\bar J_\nu(t)$ satisfies the following equation:
\begin{equation}
        {\partial \bar J_\nu\over\partial t} -
        H\left(\nu{\partial \bar J_\nu\over\partial\nu}-3\bar J_\nu\right) =
        - \bar k_\nu \bar J_\nu + \bar S_\nu,
        \label{eq:back}
\end{equation}
where, by definition, $\bar S_\nu \equiv \langle S_\nu \rangle_V$, and
$\bar k_\nu \equiv \langle k_\nu J_\nu\rangle_V / \bar J_\nu$.  In
general, $\bar k_\nu$ is not a space average of $k_\nu$, since it is
weighted by the local value of the specific intensity $J_\nu$. In the
limit of the mean free path of radiation at frequency $\nu$ being much
larger than a characteristic scale one is interested in (in our case the
size of a computational box), $\bar k_\nu= \langle k_\nu \rangle_V$.
This limit is approached for the \H2 photodissociating radiation in
the Lyman-Werner bands. 

If we rewrite equation~(\ref{eq:back}) in terms of the dimensionless
comoving photon number density $n_\nu = (aL_{box})^3 4\pi \bar J_\nu/
h_p$ where $L_{box}$ is a spatial scale (in this paper we take it
to be the comoving size of the computational box), and $h_p$ is the Planck
constant, using substitutions $dt= (a^2/H_0)d\tau$ and $\xi =
\ln(\nu)$, equation~(\ref{eq:back}) can be reduced to the following
dimensionless equation:
\begin{equation}
  {\partial n_{\xi} \over \partial \tau} = {1 \over a}{d a \over d \tau} {\partial
  n_{\xi} \over \partial \xi} - \alpha_{\xi}
  n_{\xi} + S_{\xi},
\label{eq:back2}
\end{equation}
where $\alpha_\xi = a^2 \bar k_\nu/H_0$ and $S_\xi = a^5L_{box}^3
4\pi \bar S_\nu/h_p H_0$.  Using a comoving logarithmic frequency variable,
\begin{equation}
  \bar \xi = \xi + \ln(a),
\end{equation}
equation~(\ref{eq:back2}) can be reduced to
\begin{equation}
  {\partial n_{\bar \xi} \over \partial \tau} = - \alpha_{\bar \xi}
  n_{\bar \xi} + S_{\bar \xi},
\end{equation}
which has the formal solution,
\begin{equation}
  n_{\bar \xi}(\tau+\Delta \tau) = n_{\bar \xi}(\tau)\exp\left[- 
    \int_\tau^{\tau+\Delta
      \tau} dt^\prime \alpha_{\bar
      \xi}(t^\prime)\right] + \int_\tau^{\tau+\Delta \tau}dt^\prime
  S_{\bar \xi}(t^\prime)\exp\left[-\int_{t^\prime}^{\tau+\Delta \tau} 
    d\tau^\prime \alpha_{\bar
      \xi}(\tau^\prime)\right].
\label{eq:formal}
\end{equation}
We calculate equation~(\ref{eq:formal}) at each time step, $\Delta
\tau$, of the simulation.  The two integrals inside the square
brackets on the right side of equation~(\ref{eq:formal}) can be solved
analytically. We solve the third integral numerically.  The analytical
calculations are analogous to the ones presented by
\cite{RicottiGS:01} [see \S~5 in that paper; in particular
eqs.~(28)--(31) and Fig.~11]. By definition, the line absorption
coefficient is
\[
\alpha_{\bar \xi}= {a^2c \over H_0}{\pi e^2 \over m_e c}\left[\sum_{i=LW}
  f_{osc,i}^{LW}(1-f_{i,\nu^{\prime\prime}=0}) \langle n_{H_2} \phi(\bar \xi
  ,\xi_i)\rangle_V + \sum_{i=Ly\beta}^{Lyc} f_{osc,i}^H
  \langle n_{\HI} \phi(\bar \xi,\xi_i)\rangle_V \right],
\]
where $\phi$ are the line profiles, $f_{osc,i}^{LW}$ and $f_{osc,i}^H$
are the oscillator strengths of the lines in the \H2 Lyman-Werner
bands and \HI Lyman series lines, respectively, and
$f_{i,\nu^{\prime\prime}=0}$ is the probability for the $i^{th}$ line,
calculated from \cite*{BlackD:76}, to decay to the ground vibrational
level of the $X^1\Sigma_g^+$ ground electronic state of \H2. A
fraction $f_{i,\nu^{\prime\prime}=0}$ of photons are re-emitted at the
absorption frequency; therefore, these photons are not removed from
the dissociating background.

We assume that the absorption lines in Lyman-Werner bands have
Gaussian line profiles, $\phi_G(\nu,\nu_i)=(1/\sqrt{\pi}\Delta\nu_{i})
\exp[-(\nu-\nu_i)^2/\Delta\nu_{i}^2]$, where $\Delta\nu_{i}=3\times
10^{-7} \nu_i T_{IGM}^{1/2}$ is the Doppler width of the line $i$. The
strong resonant lines in the hydrogen Lyman series typically have
Lorentzian profiles
$\phi_L(\nu,\nu_i)=(\Gamma_i/2\pi)/[(\nu-\nu_i)^2+(\Gamma_i/2)^2]$,
where $\Gamma_i=\gamma_i/2\pi$ and $\gamma_i=\sum_l A(i\rightarrow l)$
is the natural width of the $i^{th}$ \HI line. Carrying out the
integration, we have:
\begin{equation}
\begin{split}
  \int_{\tau^*}^{\tau+\Delta \tau} dt^\prime \alpha_{\bar \xi}(t^\prime) = 
  & {\pi e^2 \over m_e c}{c \over H(z)} \biggl[ \langle n_{H_2} \rangle_V
  \sum_{i=LW}{f_{osc,i}^{LW} \over
    \nu_i}(1-f_{i,\nu^{\prime\prime}=0})\Phi_G(\tau^*,\Delta \tau) + \\
  + & \langle n_{\HI} \rangle_V \sum_{i=Ly\beta}^{Lyc} {f_{osc,i}^H \over
    \nu_i}\Phi_L(\tau^*,\Delta \tau) \biggr]
\end{split}
\label{eq:1st}
\end{equation}
for the lines in the \H2 Lyman-Werner bands and the \HI Lyman lines
from Ly$\beta$ upward. Note that Ly$\alpha$ is not in the \H2
dissociating band. The functions $\Phi_G$ and $\Phi_L$ are given by
\begin{eqnarray}
  \Phi_G(\nu^\prime) &=& 
  {1 \over 2}\left[\text{erf}\left({\nu^\prime-\nu_i
        \over \Delta\nu_{i}}\right)-\text{erf}\left({\nu^{\prime\prime}-\nu_i \over
        \Delta\nu_{i}}\right)\right],\label{eq:gauss} \\
  \Phi_L(\nu^\prime) &=& 
  {1 \over \pi}\left[\arctan\left({\nu^\prime-\nu_i
        \over \Gamma_i}\right)-\arctan\left({\nu^{\prime\prime}-\nu_i \over
        \Gamma_i}\right)\right].
\label{eq:loren}
\end{eqnarray}
Here, $\nu^\prime=\nu[1+ H(\tau) \Delta \tau]$,
$\nu^{\prime\prime}=\nu[1- H(\tau) (\tau-\tau^*)]$, and
$\text{erf}(x)$ is the error function. The temperature dependence of
the Gaussian line profiles can safely be neglected. As shown in
Figure~11 of \cite{RicottiGS:01}, the integrated line profile $\Phi_G$
is almost independent of the gas temperature. In the limit of
$T_{IGM}=0$, the integrated line profile is the Heaviside function,
\begin{equation}
\lim_{T_{IGM} \rightarrow 0} \Phi_G(\nu^\prime) =
\begin{cases}
1 & \text{if ($\nu_i-\Delta_i \le \nu \le \nu_i-\Delta_{i1}$)}\\
0 & \text{if ($\nu < \nu_i-\Delta_i$) and ($\nu > \nu_i-\Delta_{i1}$)}.
\end{cases}
\end{equation}

The same method is used to calculate the \H2 dissociation cross section
$\sigma_{dis}$ from the two-step Solomon process: 
\begin{equation}
 \sigma_{dis}={\pi e^2 \over m_e c}{c \over H(\tau)}\sum_{i=LW}{f_{osc,i}^{LW}f_{dis,i} \over
  \nu_i}(1-f_{i,\nu^{\prime\prime}=0})\Phi_G(\Delta \tau),
\label{eq:2nd}
\end{equation}
where $f_{dis,i}$ is the fraction of absorptions in the \H2 line $i$
that cascade to the dissociating continuum.  In the top panel of
Figure~\ref{fig:plot} we show the first and the second terms on the
right-hand side of equation~(\ref{eq:formal}). In the bottom panel we
show the dissociation cross section $\sigma_{dis}$ (dashed line) and
the source term $S_\nu$ (solid line). In this figure we assume $\Delta
t=1$ Myr, $z=30$, and $x_{H_2}=10^{-6}$. The spectral resolution is
$\Delta \nu /\nu= 9.26 \times 10^{-6}$.

\def\capfigb{The three functions (in dimensionless units, relative to
  one) used to solve line radiative transfer in the \H2 Lyman-Werner
  bands and \HI Lyman series. Top panel: first (solid line) and second
  (dashed-line) terms on the right hand side of
  equation~(\ref{eq:formal}). Bottom panel: the source term (solid
  line) and \H2 dissociation cross section $\sigma_{dis}$ (dashed
  line). The spectral resolution is $\Delta \nu /\nu= 9.26 \times
  10^{-6}$ (\ie, 20,000 logarithmic bins in the energy range
  $11.2-13.6$ eV).}
\placefig{
\begin{figure*}[thp]
\plotone{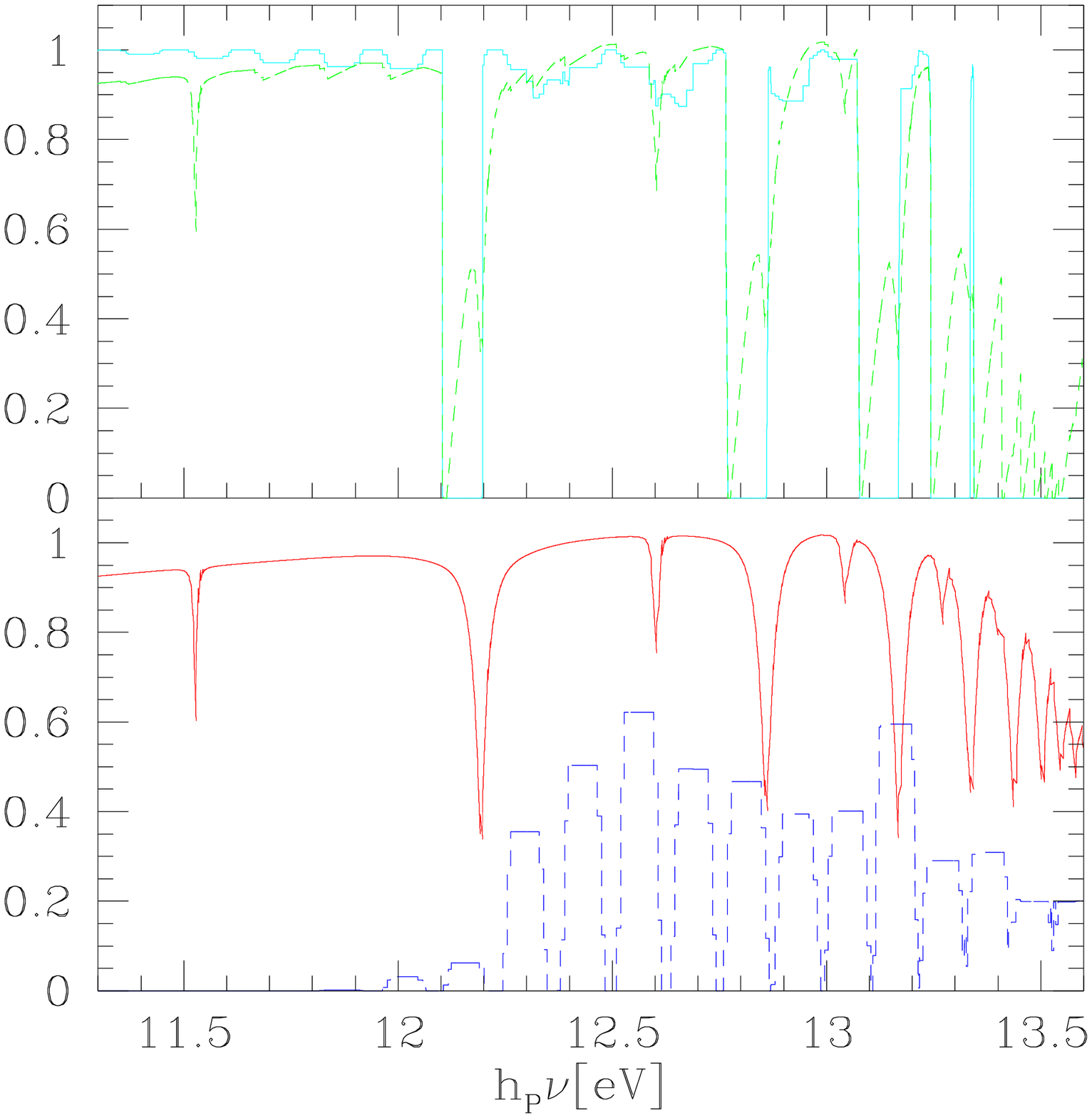}
\caption{\label{fig:plot}\capfigb}
\end{figure*}
}

\subsection{H$_2$ Chemistry and Cooling/Heating Processes}

In Figure~\ref{fig:rates} we compare the chemical rates from the three
compilations used most widely in cosmology. The labels on the top of
each panel indicate the reaction and the name of the rate coefficient
according to the author of the compilation: K \citep*{ShapiroKang:87},
A \citep*{Abel:97} and H \citep*{Galli:98}. The bottom-right panel
compares the ro-vibrational \H2 cooling functions from
\cite{Tegmark:97}, \cite*{Lepp:83}, and \cite{Galli:98}. We have
assumed $x_{H_2}=10^{-3}$; for comparison we also show the Ly$\alpha$
cooling. The thin lines show additional cooling from the reactions k7,
k9, k16, and k17 according to \cite{ShapiroKang:87}, assuming
$x_e=0.5, x_{H_2^+}=10^{-7}$ and $x_{H^-}=10^{-7}$. These cooling
processes are generally negligible. The cooling from reactions A8 and
A10, $\Lambda_{H_2} =n_H(3.53k_8 n_{H^-} + 1.38k_{10}n_{H_2^+})$ eV
cm\mmm s\m \citep{Abel:97}, can be important (horizontal dash-dotted
line).  The code can use each one of these rates, but in the following
we will use the rates and \H2 cooling function from \cite{Galli:98}.
We also include the additional cooling/heating processes from the
aforementioned reactions.  We neglect D and Li reactions and the
formation of HeH$^+$ molecules.  They are negligible in the range of
densities resolved by our simulations.

\def\capfigc{ Comparison of rate compilations found in literature:
  \cite{ShapiroKang:87} dashed lines, \cite{Galli:98} dash-dotted
  lines, and \cite{Abel:97} solid lines. The bottom-right panel shows
  H$_2$ cooling functions from \cite{Tegmark:97} (solid line),
  \cite{Lepp:83} (dashed line), and \cite{Galli:98} (dot-dashed line).
  The thin lines in the same panel show cooling rates from the
  reactions, k7, k9, k16, k17 \citep{ShapiroKang:87} and A8, A10
  \citep{Abel:97}. In the simulations shown in this paper, we adopt
  the rates and the H$_2$ cooling function from \cite{Galli:98}. We
  also include cooling from the aforementioned reactions.}
\placefig{
\begin{figure*}[thp]
\plotone{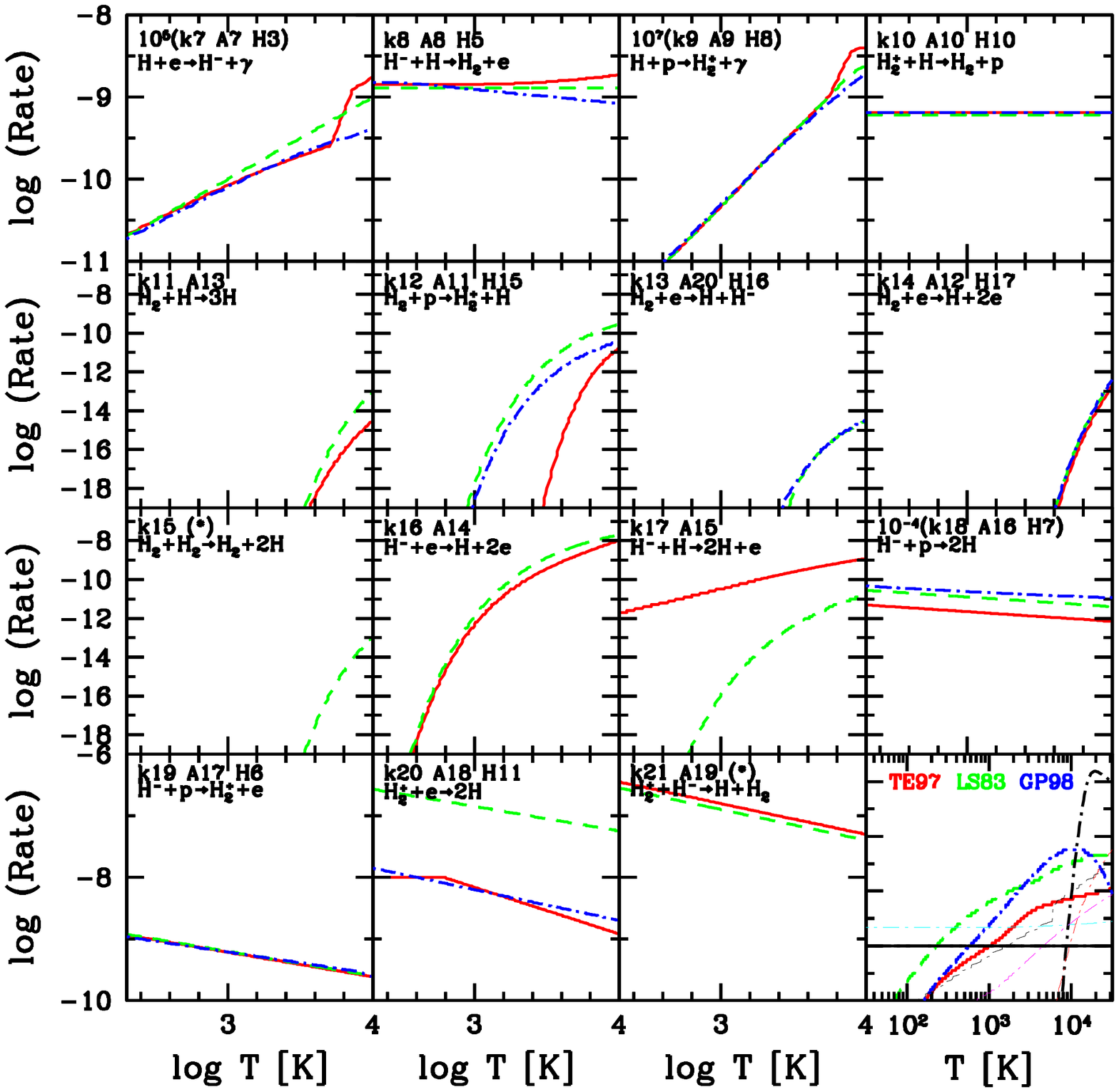}
\caption{\label{fig:rates}\capfigc}
\end{figure*}
}

\subsubsection{Heating by resonant scattering of Ly$\alpha$ photons}

Generally, in scattering processes of photons with particles, the
recoil of the particles causes the radiation field to exchange energy
with the matter. In the case of resonance lines, contrary to electron
scattering, the cumulative effects of atom recoils remain small even
for large line optical depths. This happens because a photon in the
red wing of a line is strongly biased to scatter back to the blue.
Thus, the background intensity at the resonance line frequency
develops only a slight asymmetry. The average relative shift in a
Ly$\alpha$ photon energy, $E$, after having been scattered by a
hydrogen atom at rest is $\langle \Delta E / E \rangle = -(h_p\nu/ m_H
c^2) \approx -10^{-8}$, where $m_H$ is the mass of the hydrogen atom
\citep*{Madau:97} and $h_p$ is the Planck constant.  If $T \ll h_p
\nu_\alpha /k= 1.2 \times 10^5$~K, the energy is transferred from
Ly$\alpha$ photon to the gas at a rate,
\begin{equation}
{dE_\alpha \over dt} = - \left \langle{\Delta E \over E}\right\rangle
h_p \nu_\alpha P_\alpha,
\end{equation}
where,
\begin{equation}
P_\alpha = \int d\Omega \int {J_\nu \over h_p \nu}\sigma_\nu^{Ly\alpha} d\nu,
\end{equation}
is the Ly$\alpha$ scattering rate per H atom and $\sigma_\nu^{Ly\alpha}$ the
Ly$\alpha$ absorption cross section.

The background at the Ly$\alpha$ frequency is produced mainly by
redshifted non-ionizing UV photons (\ie, the photons emitted at the
rest frame in the Lyman-Werner bands). All the other sources of
Ly$\alpha$ photons have a negligible effect on heating of neutral IGM.
For instance, compared to X-ray heating discussed in the next section,
the heating produced by the scattering of Ly$\alpha$ photons emitted
in the rest frame (not redshifted) from normal galaxies, is about 100
times less efficient \citep{Madau:97}.

\subsubsection{Secondary Ionization and Heating from X-rays}

Photoionization of \HI, \GI, and \GII by X-rays and EUV photons
produces energetic photoelectrons that can excite and ionize atoms
before their energy is thermalized. This effect can be important
\citep*{Venkatesan:01} before reionization, when the gas is almost
neutral and the spectrum of the background radiation is hard due to
the large optical depth of the IGM to UV photons.

We provide analytic fits to the Monte Carlo results of
\cite*{ShullVan:85}.  Collisional ionization and excitation of \GII by
primary electrons are neglected since, in a predominantly neutral
medium, they are unimportant.  The primary ionization rate for the
species $i=\HI, \GI,\GII$ are,
\begin{equation}
\zeta^i= 4\pi \int_{\nu_i}^\infty {J_\nu \over h_p \nu}\exp(-\tau_\nu)\sigma_\nu^i d\nu,
\end{equation}
where $\tau_\nu$ is the continuum optical depth, and $\sigma_\nu^i$ is the
photoionization cross section of the species $i$.
Secondary ionizations enhance the photoionization rates as follows:
\begin{eqnarray}
\zeta_s^{\HI} = \zeta^{\HI}+\sum_{i=\HI,\GI,\GII} \zeta^i\langle
\Phi^{\HI}(E_0^i,x_e)\rangle \label{eq:z1}\\
\zeta_s^{\GI} = \zeta^{\GI}+\sum_{i=\HI,\GI,\GII} \zeta^i\langle
\Phi^{\GI}(E_0^i,x_e)\rangle,
\end{eqnarray}
where $\langle \Phi^{\HI}(E_0^i,x_e)\rangle$ and $\langle
\Phi^{\GI}(E_0^i,x_e)\rangle$ express the average number of secondary
ionizations per primary electron of energy $E_0^i = h_p \nu - I^i$
weighted by the function $(J_\nu/h_p \nu) \exp(-\tau_\nu)\sigma_\nu^i$.
Here $I^i=h_p \nu_i$ is the ionization potential for the species $i$.

The photoionization heating rates for the species $i=\HI, \GI$ are given by,
\begin{equation}
\Gamma^i= 4\pi \int_{\nu_i}^\infty {J_\nu \over h_p \nu}\exp(-\tau)\sigma_\nu^i
E_h(E_0^i,x_e) d\nu.\label{eq:heat}
\end{equation}
In Appendix~\ref{ap:sec} we give analytic fits to the functions
$\Phi^{\HI}, \Phi^{\GI}$ and $E_h$. We express these functions in a
form suited to minimize the computational time for the integrals in
equations~(\ref{eq:z1})--(\ref{eq:heat}).

\subsection{Sources spectrum and \fesc \label{ssec:fesc}}

In \cite{RicottiGS:01} we showed that the SED of a galaxy
and \fesc are the main parameters that determine the
relevance of {\em positive feedback} effects. In particular, we found
that the \pop3 (metal-free) and mini-quasar (with spectral
index $\alpha = 1.8$) SED produce similar positive feedback regions
(PFR). In this section, we explain how these ingredients are included
in the code. We do not use the mini-quasar SED in our simulations.
Apart from producing a stronger X-ray background, their effects should
be analogous to those of the \pop3 SED.

Resolution elements that sink below the resolution limit of the
simulation are allowed to form stars and therefore become sources of
radiation according to the following equations:
\begin{eqnarray}
  {d\rho_* \over dt} &=& \epsilon_* {\rho_g \over t_*}, \label{eq:sf}\\
  S_\nu &=& {\epsilon_{UV} \over 4\pi} c^3{d\rho_* \over dt}
  g_\nu e^{-\tau_\nu}.
\end{eqnarray}
Star formation is implemented using the Schmidt law, where
$\epsilon_*$ is the star formation efficiency, $\rho_*$ and $\rho_g$
are the stellar and gas density, respectively, and $t_*$ is the
maximum of the dynamical time and cooling time.  The parameter
$\epsilon_{UV}$ is the ratio of energy density of the ionizing
radiation field to the gas rest-mass energy density converted into
stars, $g_\nu$ [photons Hz$^{-1}$] is the normalized SED, 
\[
\int_{\nu_0}^\infty g_\nu d\nu =1,
\]
and $\tau_\nu$ is the internal opacity of the source. The escape
fraction of ionizing photons, \fesc, is defined as:
\begin{equation}
\langle f_{esc} \rangle =\int_{\nu_0}^\infty g_\nu e^{-\tau_\nu} d\nu,
\label{eq:f1}
\end{equation}
where $\nu_0$ is the \HI Lyc frequency. A value of \fesc$<1$ has the
additional effect of producing harder source spectra because the gas
optical depth is higher for UV and FUV photons than for X-ray photons.
We include this effect modifying the SED using a frequency dependent
optical depth
\begin{equation}
\tau_\nu = N_{\HI}(\sigma_{\HI}+a_0\sigma_{\GI}+a_1\sigma_{\GII}).
\label{eq:f}
\end{equation}
Here, $a_0=N_{\GI}/N_{\HI}$ and $a_1=N_{\GII}/N_{\HI}$ where $N_i$ is
the column density of the species/ion $i$. The unknown parameters,
$a_0$ and $a_1$, depend on the interstellar medium (ISM) properties of
the galaxies, including clumping. Based on the results of simple 1D
radiative transfer simulations presented in \cite{RicottiGS:01}, we
chose $a_0=0.01$ and $a_1=10$ when we use the \pop3 SED, and $a_0=0.1$
and $a_1=10$ when we use the \popII SED.  These values are calculated
placing the source of radiation in the center of a spherical galaxy
with a gas density profile calculated with high-resolution numerical
simulations \citep*{Abel:00}. We did not consider the effects of a
clumpy ISM.  Given the value of \fesc, $N_{\HI}$ is calculated from
equations~(\ref{eq:f1})--(\ref{eq:f}) using a bisection algorithm.

The spatial resolution of the code is determined by the dimensionless
SLH parameter, $B_*$.  The code switches from Lagrangian to Eulerian
when the deformation of the spatial grid exceeds a critical value that
is approximatively $1/B_*$ the undeformed cell size. Therefore, the
comoving spatial resolution is about $L_{box}/(N_{box}B_*)$,
where $L_{box}$ is the box size and $N_{box}^3$ is the number of cells
in the box.  Usually we use $B_*=10$ for $N_{box}=64$ ($262,144$
cells), $B_*=16$ for $N_{box}=128$ (2 million cells) and $B_*=25$ for
$N_{box}=256$ (17 million cells). The gravity force is calculated by a
P$^3$M algorithm with the Plummer softening parameter equal to
$1/2B_*$. This ensures that the gravity force is accurate on our
resolution scale, and thus $1/B_*$ is a real rather than nominal
resolution. The nominal resolution is twice higher.

We allow the spectral energy distribution, $g_\nu$, to be chosen 
from the following:\\
i) \popII: metallicity $Z = 0.04$ Z$_\odot$, evolutionary tracks
evolved to $t = 1$ Gyr, continuous SF law, Salpeter IMF with star
masses between $1<M_*<100$~M$_\odot$ \citep*{Leitherer:95}. Wolf-Rayet
stars are
responsible for the substantial EUV emission in this SED.\\
ii) \pop3 : metallicity $Z=0$, non-evolved, instantaneous SF law,
Salpeter IMF with star masses between $1<M_*<100$~M$_\odot$
\citep*{Tumlinson:00}. Note the importance of
\GII ionizing radiation in this SED.\\
iii) \pop3 : metallicity $Z=0$, evolutionary tracks evolved to $t=1$
Gyr, continuous SF law, Salpeter IMF with star masses between
$1<M_*<100$~M$_\odot$ \citep*{Brocato:99, Brocato:00}. This SED
calculation differs substantially from (ii) since it does
not include \GII ionizing photons.\\
In Figure~\ref{fig:pop3} we show the analytical fits implemented in
the code for these three SEDs. In our simulations we will use only the
non-evolved \pop3 SED (ii) and the \popII SED (i). The evolved \pop3
SED (iii) is implemented in the code but never used. We show it just
for comparison since calculations from different authors of the \pop3
SEDs are still in disagreement.

\def\capfigd{ Analytical fits to the normalized \popII SED for
  a continuous SF law evolved to $t=1$ Gyr (solid line), non-evolved
  \pop3 SED for an instantaneous SF law (short-dashed line) and \pop3 SED
  for a continuous SF law evolved to $t=1$ Gyr (long-dashed line). See
  text for further explanation.}
\placefig{
\begin{figure*}[thp]
\plotone{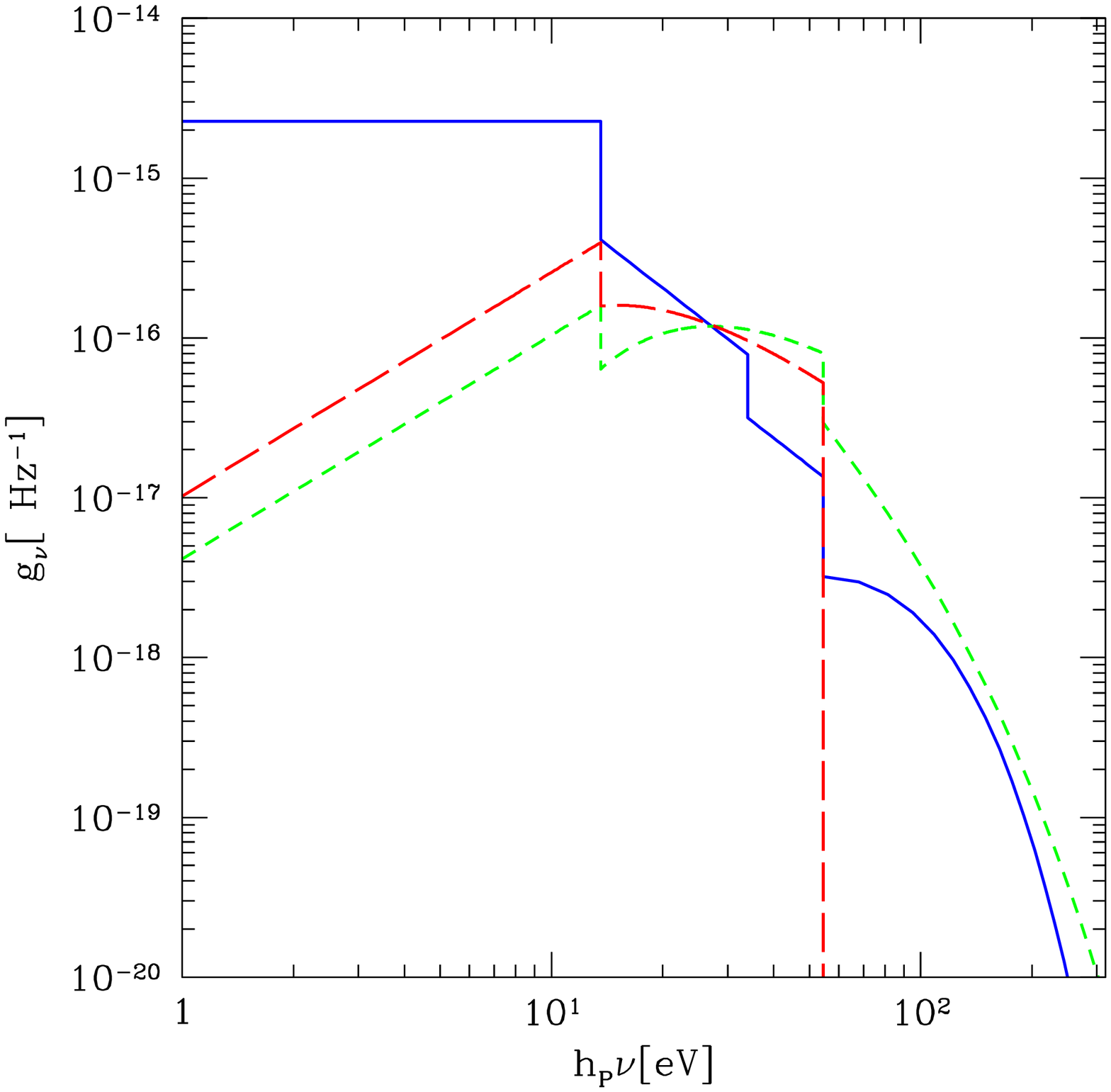}
\caption{\label{fig:pop3}\capfigd}
\end{figure*}
}

The free parameter $\epsilon_{UV}$ is the ratio of energy density of
the ionizing radiation field to the gas rest-mass energy density
converted into stars ($\rho_* c^2$). Thus, $\epsilon_{UV}
=(\overline{h_p \nu} / m_Hc^2) \epsilon_{UV}^*$, where
$\epsilon_{UV}^*$ is the number of ionizing photons per baryon
converted into stars. The value of $\epsilon_{UV}$ depends on the IMF
and the metallicity of the stellar population.  Assuming a Salpeter
IMF with star masses between $1~{\rm M}_\odot \le {\rm M}^* \le
100~{\rm M}_\odot$, we have $(\epsilon_{UV}/4\pi)=1.1 \times 10^{-5}$,
$\overline{h_p \nu} / (13.6~{\rm eV})=1.76$ for the \popII SED and
$(\epsilon_{UV}/4\pi)=2.5 \times 10^{-5}$, $\overline{h_p \nu} / (13.6~{\rm
  eV}) =2.47$ for the (ii) \pop3 SED. Theoretical works
\citep*{Uehara:96, Larson:98, Nakamura:99,Nakamura:01} have shown that
the IMF of metal-free stars could be dominated by massive stars or
might have a bimodal IMF (lacking in intermediate mass stars).  If the
IMF is dominated by very massive stars ($M \sim 300$ M$_\odot$) that
eventually evolve into black holes \citep*{Bromm:01}, it is possible
to have $(\epsilon_{UV}/4\pi) \gg 2.5 \times 10^{-5}$.

The free parameter \fesc is the fraction of ionizing photons escaping
from the resolution element of the simulation. According to its
definition, this parameter is, in general, resolution and time
dependent. Theoretical studies \citep*{RicottiS:00, Wood:00} have
shown that \fesc, defined as the fraction of ionizing photons escaping
from a galaxy halo\footnote{The comoving radius at 200 times the
  background density of a DM halo with mass $M_{DM}$ that virializes
  at redshift $z_{vir}$ is $r_{200} \approx (2.4~h^{-1}{\rm kpc})(M_{DM}/10^6
  h^{-1}{\rm M}_\odot)^{1/3}[(1+z)/(1+z_{vir})]$.}, decreases at
increasing redshift and increasing mass of the halos (assuming
constant star formation efficiency). In particular, \cite{RicottiS:00}
provide analytic formulae for \fesc$(z_{vir},\epsilon,f_g,M_{DM})$ as
a function of the star formation efficiency (SFE), $\epsilon$,
normalized to the Milky Way value, the redshift of virialization,
$z_{vir}$, the DM halo mass, $M_{DM}$, and the fraction of collapsed
gas $f_g$. The following formula, for example,
\begin{equation}
\langle f_{esc} \rangle = {4.44 \over (\epsilon f_g)^{1/3}}\left({10^6
   ~{\rm M}_\odot \over M_{DM}}\right)^{1 \over
   3}\exp \left[-0.94(1+\log f_g^{2 \over 3}){z_{vir}+1 \over
   \epsilon^{1/3}}\right],
\label{eq:fesc}
\end{equation}
is valid for \p3 objects ($M_{DM} \le 10^8 M_\odot [(1+z_{vir})
/10]^{-1.5}$), a constant SFE $4<\epsilon<400$ (\ie, the ionizing flux
is proportional to the gas mass of the halo) and assuming a power law
for the ionizing photon luminosity function of the OB associations
with slope $\alpha=2$ \citep*{Dove:94, Kennicutt:89}, lower limit
$S_1=10^{48}$ photons s\m, and no upper limit.
Equation~(\ref{eq:fesc}) cannot be used in the code, because the
definition of \fesc depends on the spatial resolution of the
simulation. Instead, \fesc in equation~(\ref{eq:fesc}) is a function
of the halo mass.  Nevertheless, even if \fesc in the code is larger
than \fesc in equation~(\ref{eq:fesc}), depending on the spatial
resolution of the code, the functional dependence of \fesc on the
virialization redshift should be robust. In this study, we will
consider \fesc as a constant free parameter. Equation~(\ref{eq:fesc})
is simply used to provide a first-order estimate of reasonable values
for \fesc at high redshift.  We plan to relax this assumption in
future work.

\section{Resolution Studies \label{sec:resol}}

In this section we investigate the numerical convergence of our
simulations.  Simulating the formation of the first galaxies is
computationally challenging. It requires high mass resolution in order
to resolve the internal structure of each low-mass object with a
sufficient number of dark matter particles and grid cells.  Moreover,
since the first objects arise from the collapse of the most massive
and rare density perturbations at redshift $z \sim 30$, we need a
sufficiently large box to include at least a few 3 $\sigma$
perturbations of the initial density field.  These requirements
translate into the need for simulations with large mass dynamical
range (\ie, a large number of cells and DM particles in the box). We
have been able to run a few simulations with the maximum feasible
resolution of $256^3$ cells, but the bulk of the study consists of
runs with $128^3$, and $64^3$ cells.

In Table~\ref{tab:1} we list the runs used to test the convergence of
the simulation. In all these test runs (but one: 256L1p3) the
radiative transfer algorithm has been shut down. Therefore, the stars
were able to form but did not produce any feedback effects.

\def\tabone{
\begin{deluxetable}{lcccccc}
\tablecaption{List of simulations.\label{tab:1}}
\tablewidth{0pt}
\tablehead{
\colhead{RUN} &\colhead{$N_{box}$} & \colhead{$L_{box}$} & \colhead{Mass
  Resolution} & \colhead{$B_*$} & \colhead{$\left({\epsilon_{UV}\over 4\pi}\right)$\fesc} &
\colhead{$\epsilon_*$} \\
\colhead{} & \colhead{} & \colhead{($h^{-1}$ Mpc)} & \colhead{($h^{-1}$ M$_\odot$)
} & \colhead{} & \colhead{} & \colhead{}
 }
\startdata

 64L05B8   & 64  & 0.5  & $3.94\times 10^4$ & 8  & 0 &  0.2 \\
 64L05     & 64  & 0.5  & $3.94\times 10^4$ & 10 & 0 &  0.2 \\
 64L05B12  & 64  & 0.5  & $3.94\times 10^4$ & 12 & 0 &  0.2 \\
 64L05B16  & 64  & 0.5  & $3.94\times 10^4$ & 16 & 0 &  0.2 \\
 \\
 64L1      & 64  & 1.0  & $3.15\times 10^5$ & 10 & 0 &  0.2 \\
 64L1B16   & 64  & 1.0  & $3.15\times 10^5$ & 16 & 0 &  0.2 \\
 \\
 128L05    & 128 & 0.5  & $4.93\times 10^3$ & 10 & 0 &  0.2 \\
 128L1     & 128 & 1.0  & $3.94\times 10^4$ & 10 & 0 &  0.2 \\
 128L1B16  & 128 & 1.0  & $3.94\times 10^4$ & 16 & 0 &  0.2 \\
 \\
 256L1  & 256 & 1.0  & $4.93\times 10^3$ & 10 & 0 &  0.2 \\
 256L2  & 256 & 2.0  & $3.94\times 10^4$ & 10 & 0 &  0.2 \\
 \\
 256L1p3\tablenotemark{a}  & 256 & 1.0  & $4.93\times 10^3$ & 25 & $2.5 \times 10^{-6}$ &  0.1 \\
\enddata

\tablecomments{Parameter description. {\em Numerical parameters:}
  $N_{box}^3$ is the number of grid cells, $L_{box}$ is the box size
  in comoving \Mpc, $B_*$ is a parameter that regulates the maximum
  deformation of the Lagrangian mesh: the spatial resolution is $\sim
  L_{box}/(N_{box}B_*)$. {\em Physical parameters:} $g_\nu$ is the
  normalized SED, $\epsilon_*$ is the star formation efficiency,
  $\epsilon_{UV}$ is the ratio of energy density of the ionizing
  radiation field to the gas rest-mass energy density converted into
  stars (depends on the IMF), and \fesc is the escape fraction of
  ionizing photons from the resolution element.}

\tablenotetext{a}{$g_\nu$ is the \pop3 SED, modified assuming \fesc =
  0.1, $a_0=N_{\GI}/N_{\HI}=0.01$ and $a_1=N_{\GII}/N_{\HI}=10$ where
  $N_i$ is the column density of the species/ion $i$ (see
  \S~\ref{ssec:fesc}).}

\end{deluxetable}
}
\placefig{\tabone}

\subsection{Mass resolution and Box Size}

We check the convergence of the simulations with two sets of runs. The
first set that includes runs 64L1, 128L1, and 256L1 has a constant
box size $L_{box}=1$ \Mpc, and mass resolutions $M_{DM}=3.15\times
10^5, 3.94\times 10^4$, and $4.93\times 10^3$ \Ms, respectively. The
second set that includes runs 64L05, 128L1, and 256L2 has a
constant mass resolution $M_{DM}=3.94\times 10^4$ \Ms and box
sizes $L_{box}=0.5, 1$, and 2 \Mpc, respectively. If we assume that a
physical output of the simulation, $C$, scales with $N_{box}$ as
\begin{equation}
C = C_\infty + {\Delta \over (N_{box})^\alpha}
\label{eq:conv}
\end{equation}
(a usual assumption in convergence analysis of numerical simulations),
we can determine the asymptotic limit $C_\infty= \lim_{N_{box}
  \rightarrow \infty} C$ if we have three simulations that differ only
in the value of $N_{box}$. Since the choice of
equation~(\ref{eq:conv}) is somewhat arbitrary, the estimate of the
convergence of the simulations is perhaps not rigorous and the results
should be interpreted simply as a sign of convergence. We never used
resolution studies to correct the results our the simulations for
resolution or box size effects.

The curves in Figure~\ref{fig:res} show equation~(\ref{eq:conv}) as a
function of $N_{box}$, where the observable $C$ is the redshift $z$ at
which the star fraction is $\log f_{star}=\log( \rho_{star}/\rho_b) =
{\rm const.}$ In Figure~\ref{fig:res}~(left) the data points are for
the first set of simulations; therefore, the asymptotic limits
$C_\infty$ (horizontal lines) show the redshifts at which $\log
f_{star}= -6,-5,-4,-3,-2.2$ for a simulation with infinite mass
resolution and $L_{box}=1$.  Figure~\ref{fig:res}~(right) is analogous
to Figure~\ref{fig:res}~(left) but for the second set of
simulations. Here, the asymptotic limits $C_\infty$ show the redshifts
at which $\log f_{star}= -6,-5,-4,-3.5$ for a simulation with infinite
box size and mass resolution $M_{DM}=3.94\times 10^4$ \Ms.

\def\capfige{Resolution studies. (Left) we use a set of three
  simulations (256L1, 128L1, and 64L1) with different mass resolution
  and $L_{box}=1$ \Mpc, to extrapolate to the infinite resolution limit
  (thin horizontal lines). The curves, from top to bottom, show $z+1$
  as a function of $N_{box}$ [eq.~(\ref{eq:conv})] at constant $\log
  f_{stars}=-6,-5,-4,-3,-2.2$, and crosses on each curve show the
  three data points. (Right) same as left but for a set of three
  simulations (256L2, 128L1, and 64L05) with different box sizes and
  the same mass resolution. The thin horizontal lines show the infinite
  box size limits. The lines from top to bottom are at constant $\log
  f_{stars}=-6,-5,-4,-3.5$.} 
\placefig{
\begin{figure*}[thp]
\plottwo{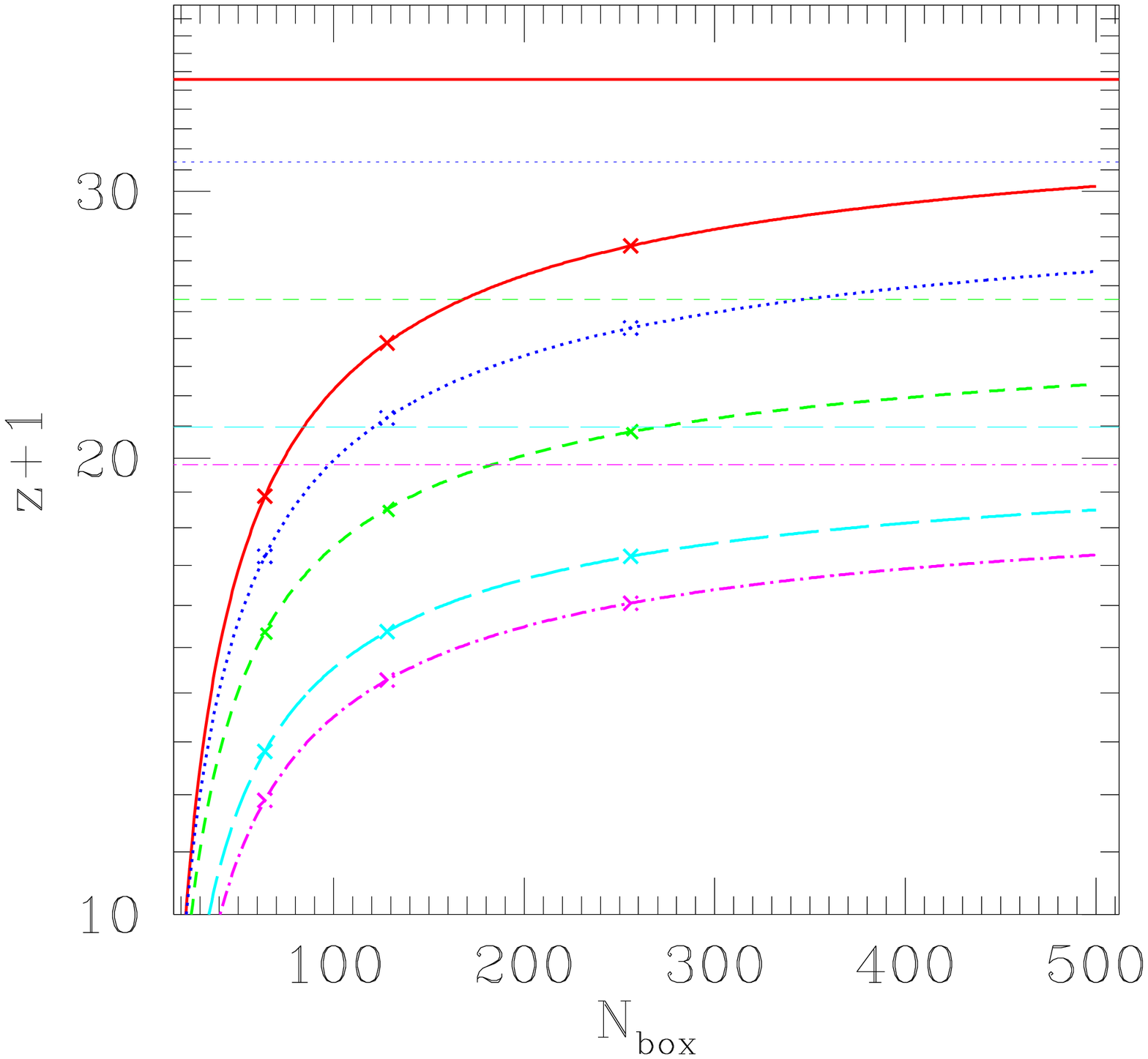}{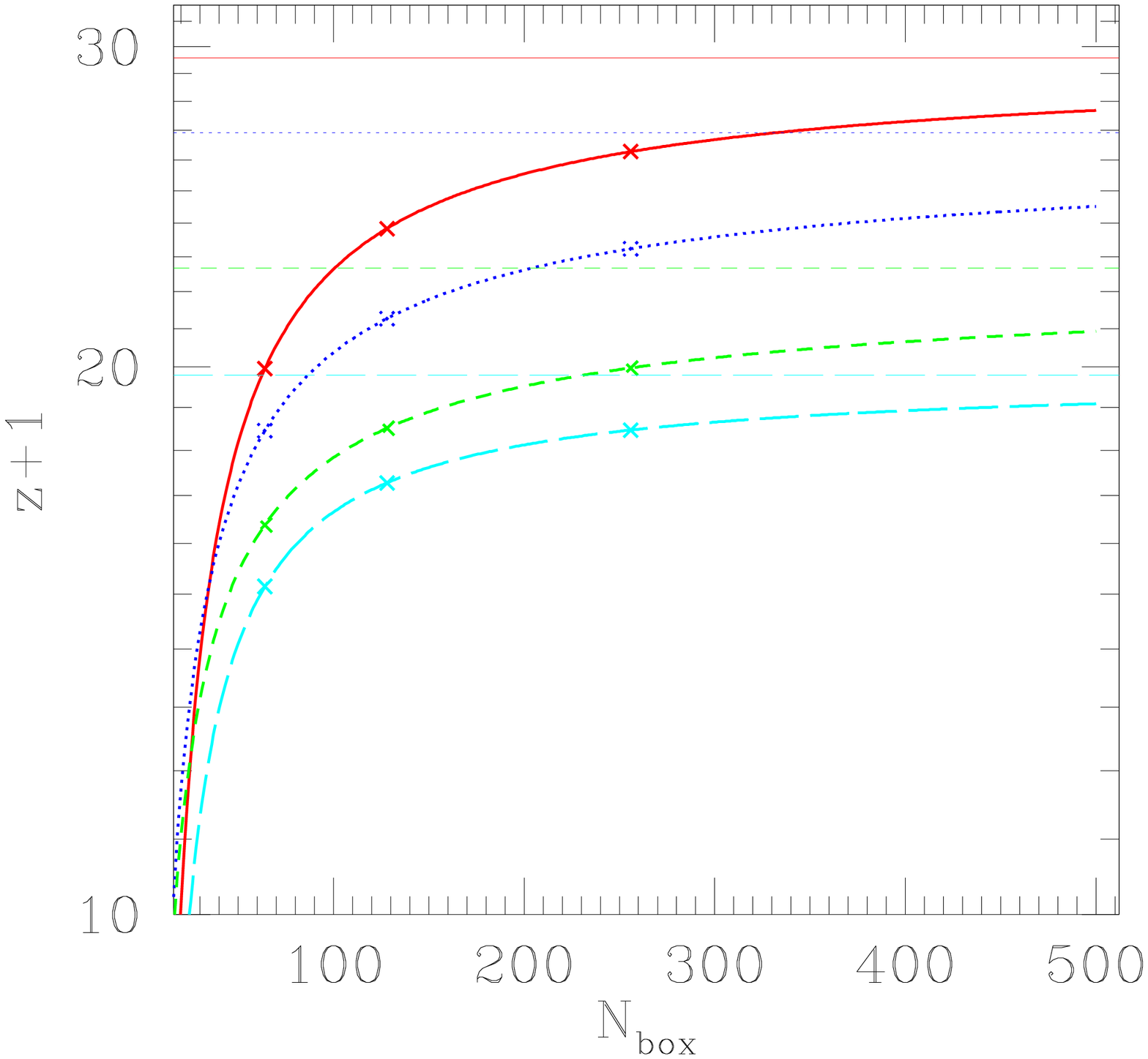}
\caption{\label{fig:res}\capfige}
\end{figure*}
} 

In Figure~\ref{fig:sfr1} we summarize the mass resolution and box-size
convergence studies. The curves show the fraction of baryons in stars,
$f_{star}$, for the two sets of simulations discussed in the previous
paragraph. The triangles show $f_{star}$ in the limit of a simulation
with infinite mass resolution and $L_{box}=1$ \Mpc. The squares show
$f_{star}$ for a simulation with infinite box size and mass resolution
$M_{DM}=3.94\times 10^4$ \Ms.  Figure~\ref{fig:sfr1} shows that the
256L1 run is the closest to the convergence limit.

\def\capfigf{Fraction of baryons in stars versus redshift for two sets
  of simulations with different resolution and box sizes. The three
  thick lines show 256L1, 128L1, and 64L1 simulations with constant
  box size ($L_{box}=1$ \Mpc), and varying mass resolutions
  ($M_{DM}=3.15\times 10^5, 3.94\times 10^4$, and $4.93\times 10^3$
  \Ms, respectively). The triangles show $f_{star}$ as a
  function of redshift in the limit of a simulation with infinite mass
  resolution and $L_{box}=1$ \Mpc. The three thin lines show 64L05,
  128L1, and 256L2 simulations, with constant mass resolution
  ($M_{DM}=3.94\times 10^4$ \Ms) and varying box sizes
  ($L_{box}=0.5, 1$, and 2 \Mpc, respectively). The squares show
  $f_{star}$ as a function of redshift in the limit of infinite box
  size and mass resolution $M_{DM}=3.94\times 10^4$ \Ms. Note
  that the simulation 128L1 appears in both sets; therefore there are
  only five lines in the plot.}
\placefig{
\begin{figure*}[thp]
\plotone{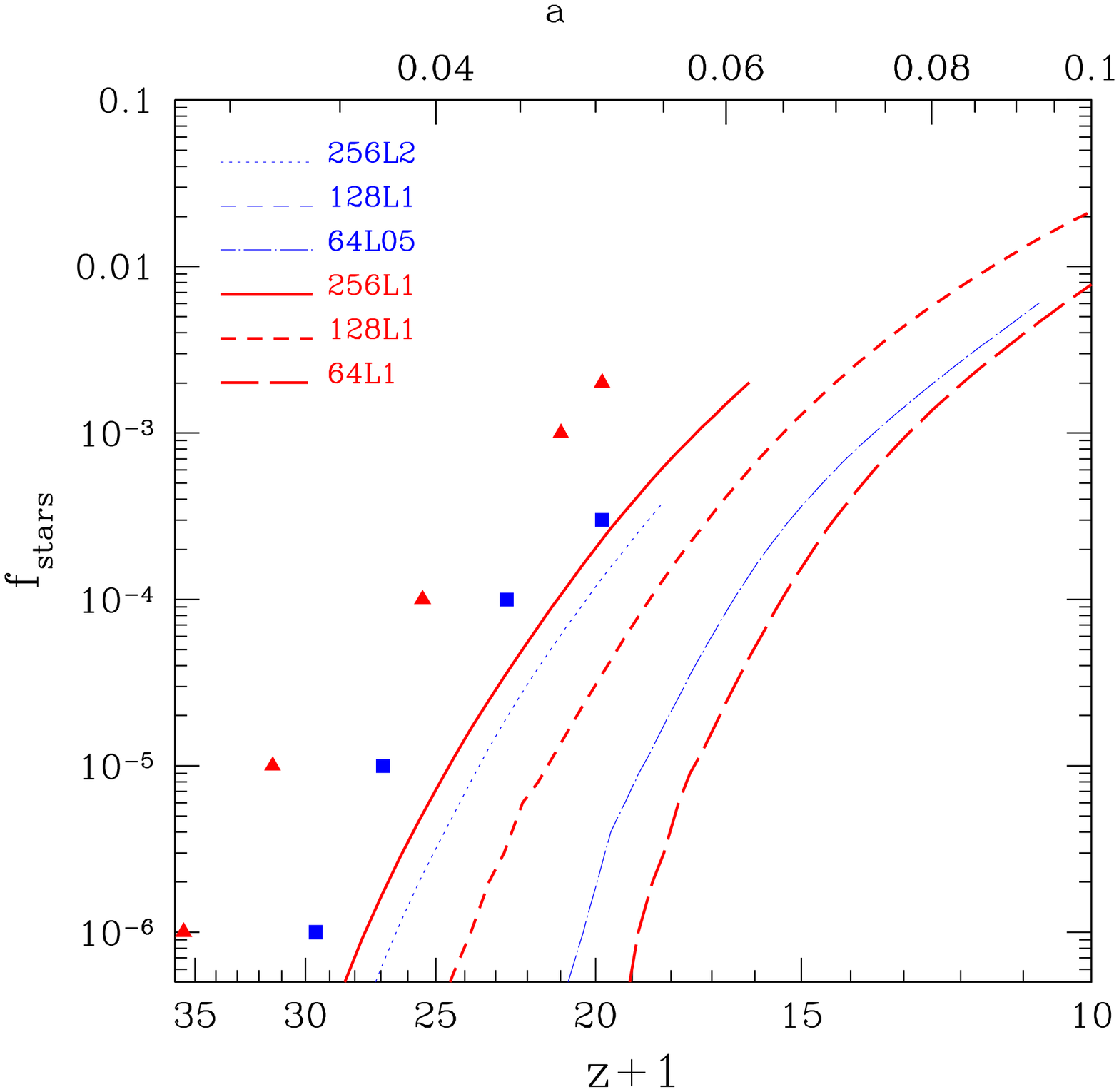}
\caption{\label{fig:sfr1}\capfigf}
\end{figure*}
}

The level of convergence of a simulation can be easily understood from
the plot in Figure~\ref{fig:colMF}. The cusp-shaped curves show the
fraction of cooled gas in collapsed objects at
$z+1=35,30,25,20,15,10,4$ (from bottom to top) calculated with the
Press-Schechter formalism. The gas collapsed in the DM halo is assumed
to be shock-heated to about its virial temperature. In order to form
stars, the gas cooling time $t_{cool}(z)$ must be shorter than the
Hubble time\footnote{For a $\Lambda$CDM cosmology with
  $(\Omega_m,\Omega_\Lambda,h)=(0.3, 0.7, 0.7)$ the age of the
  universe at $z+1 = 35, 30, 25, 20, 15, 10$ and 4, is $t_H = 82, 103,
  136, 190, 292, 537$ Myr and 2 Gyr, respectively} $1/H(z)$.  We
roughly estimate the fraction of baryons able to cool and form stars
by multiplying the virialized object fraction by the cooling
efficiency $\epsilon_{cool}=max[1/t_{cool}(z)H(z), 1]$.  Note that
$\epsilon_{cool} < 1$ for masses typical of both \p3 objects and
galaxy clusters.  The comoving size of the box determines the DM mass
of the most massive halo in the simulation, and the mass resolution
determines the mass of the least massive object resolved. The oblique
lines show the DM mass of the most massive halo in simulations of box
sizes $L_{box}=0.5, 1, 2$ \Mpc (from top to bottom). The vertical lines
show the mass of the least massive DM halo for which we resolve the SF
in simulations with mass resolution $M_{DM}=4.93 \times 10^3, 3.94
\times 10^4$, and $3.15 \times 10^5$ \Ms (from left to right).
As a quick reference, the vertical lines correspond to $256^3, 128^3,
64^3$ cubes with $L_{box}=1$ \Mpc. These minimum masses are about 100
times the mass resolution of the simulation. That means that we need
to resolve each halo with about 100 cells.

\def\capfigg{Fraction of collapsed gas as a function of DM halo mass
  at $z+1=35,30,25,20,15,10,4$ (from bottom to top).  We use the
  Press-Schechter formalism and a simple estimate for the fraction of
  collapsed gas based on the cooling time (see text). The thin
  portions of the curves correspond to virial temperatures $10^4
  \simlt T_{vir} \simlt 10^5$ K. In these halos, the cooling is very
  efficient and all the gas is collapsed; the thick portions of the
  curves on the left are \p3 objects ($T_{vir} < 10^4$ K) and on
  the right objects with $T_{vir} > 10^5$ K.  The vertical lines
  correspond to the smallest halo mass fully resolved by simulations
  with mass resolution $M_{DM}=4.93 \times 10^3, 3.94 \times 10^4$,
  and $3.15 \times 10^5$ \Ms (from left to right). As a quick
  reference they correspond to $256^3, 128^3, 64^3$ cubes with
  $L_{box}=1$ \Mpc. The oblique lines show the largest halo mass that
  we can find in cubes with $L_{box}=0.5, 1, 2$ \Mpc. For example, the
  two thicker vertical and oblique lines delimit the galaxies fully
  resolved in the 256L1 simulation.}
\placefig{
\begin{figure*}[thp]
\plotone{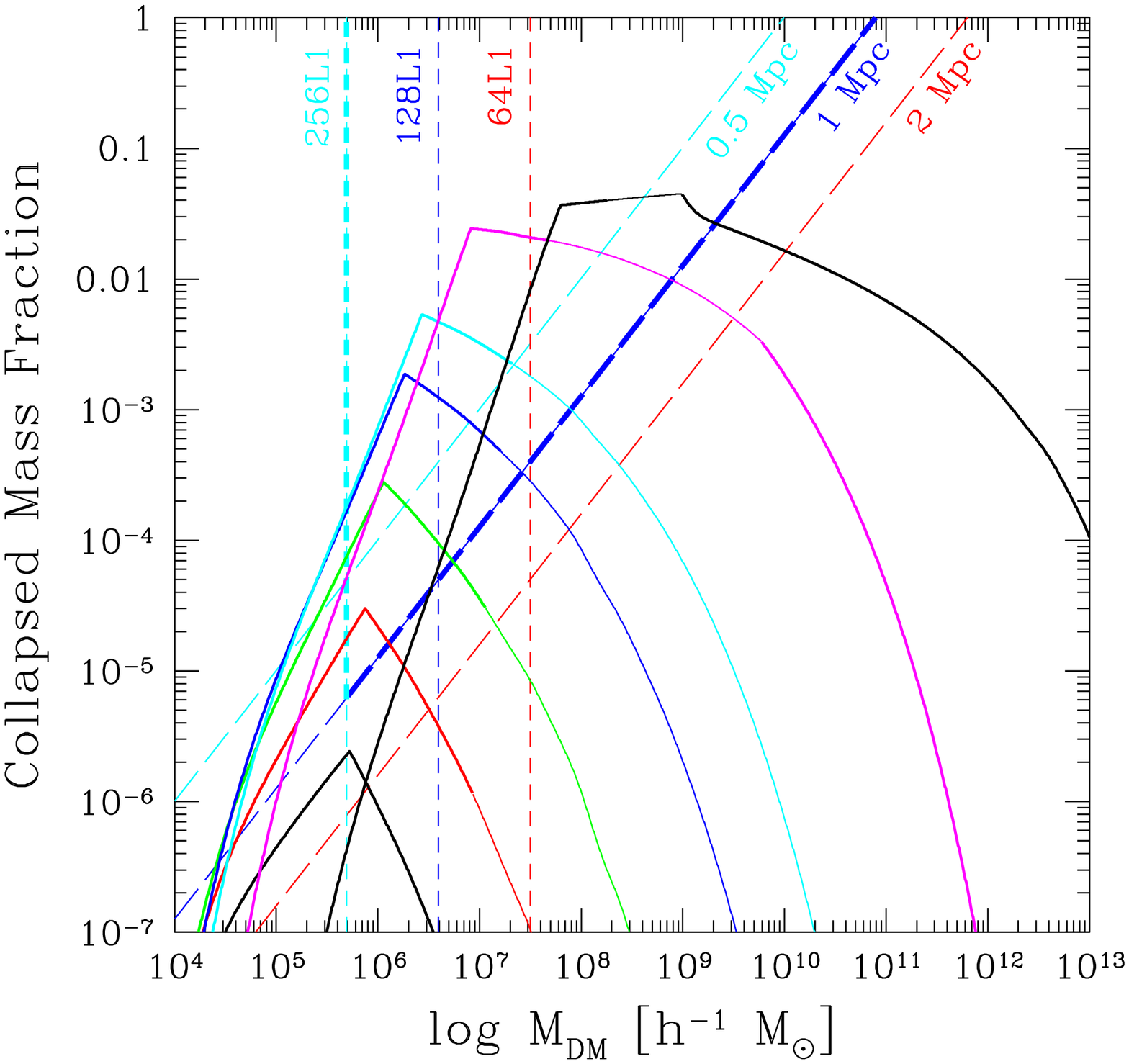}
\caption{\label{fig:colMF}\capfigg}
\end{figure*}
} 

Given the box size and the mass resolution of the simulation, the
point of intersection of the appropriate vertical and oblique lines
shows the redshift of formation and the mass of the first object in
the simulation. For example, the two thicker vertical and oblique
lines delimit the masses resolved in the $256^3$ cells, $L_{box}=1$
\Mpc simulation. In this simulation, the first objects have masses $M
\sim 5 \times 10^5$ \Ms and form between $z=29$ and $z=34$.
Moreover, this simulation can resolve the bulk of the collapsed baryons
up to redshift $z \sim 5$.

In this study we show both integrated physical quantities such as SFR,
background radiation intensity, chemical abundances, and also the
properties of individual bound objects formed in the simulations.  We
identify bound objects with the DENMAX algorithm
\citep{Bertschinger:91}. The results of DENMAX depend on the choice of
the free parameter, $G=L_{box}/L_G$, which defines the smoothing
length, $L_G$, of the particle density field. In Appendix~\ref{app:B}
we show how to select the smoothing parameter when comparing
simulations with different resolution/box sizes.

In Figure~\ref{fig:sf} we compare the SFR at $z=15$ as a function of
halo mass for simulations with different resolutions and box
sizes.  The halos have been identified using DENMAX with smoothing
parameter $G=400$ for the 128 boxes and $G=1000$ for the 256 box. It
is evident that the 128L1 run does not sufficiently resolve small mass
($M_{DM} \le 2 \times 10^7$ \Ms) halos, and the 128L0.5 run does
not have enough large mass objects.

\def\capfigh{Global SFR at $z=15$ as a function of DM halo mass. The three
  histograms show the same simulation at different resolutions and box
  sizes. The solid line shows the simulation with $256^3$ cells and
  $L_{box}=1$ \Mpc, the dashed-line $128^3$, $L_{box}=1$ \Mpc, and the
  dotted line $128^3$, $L_{box}=0.5$ \Mpc. The halos have been
  identified using DENMAX with smoothing parameter $G=400$ for the $128^3$
  boxes and $G=1000$ for the $256^3$ box. The three horizontal lines above
  the histograms show the total (integrated over all masses) SFR for the
  same simulations.}
\placefig{
\begin{figure*}[thp]
\plotone{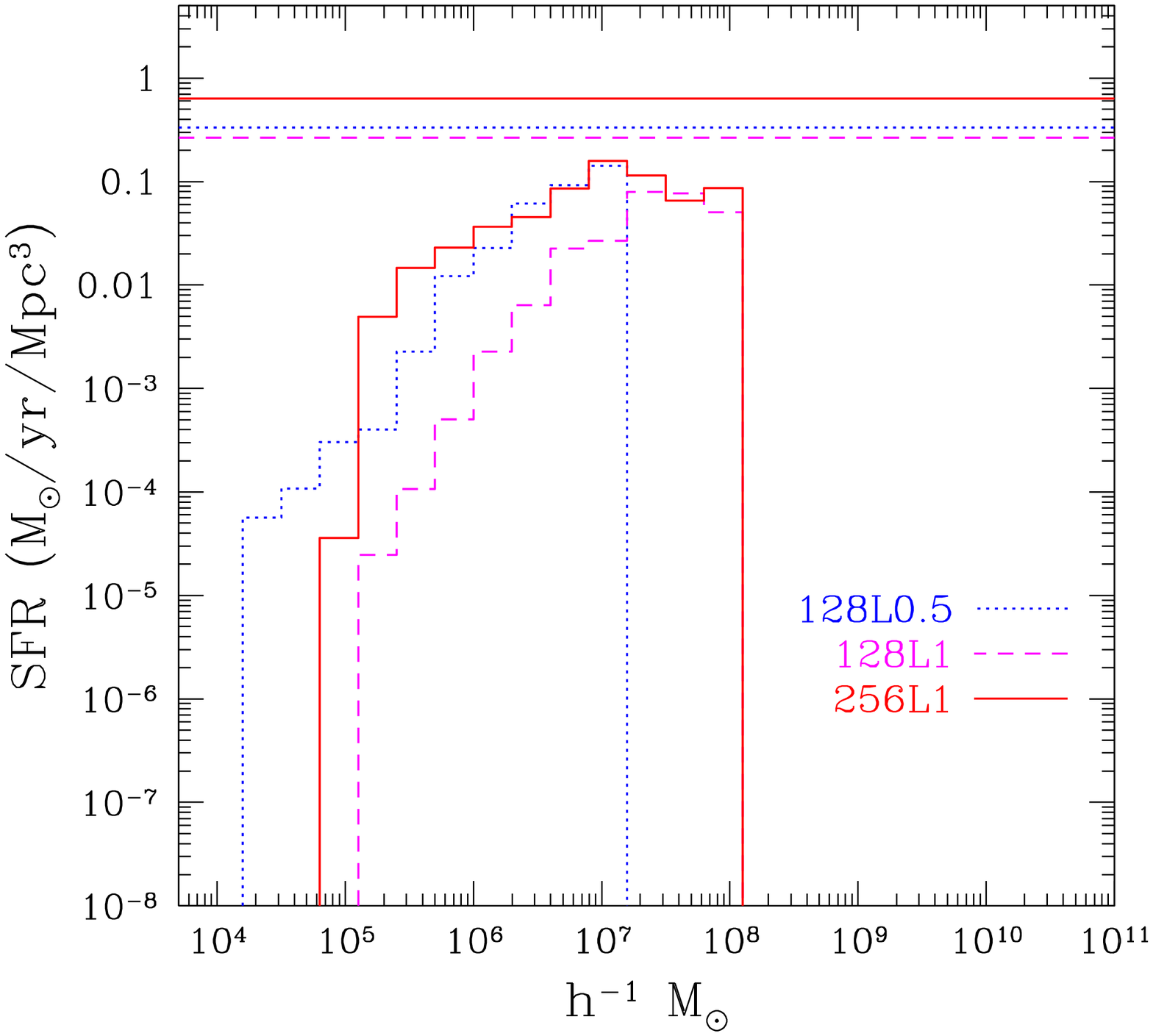}
\caption{\label{fig:sf}\capfigh}
\end{figure*}
} 

The minimum redshift to which we can evolve the simulations depends on
the box size. For 2, 1, and 0.5 \Mpc boxes, we can evolve the
simulation until redshifts $z \approx 6, 9$, and 12 respectively.
Thereafter, scales of the order of the box size become nonlinear, and
the box ceases to be a representative sample of the universe.

\subsection{Spatial resolution}

The spatial resolution of the grid is regulated by the softening
parameter $B_*$, which sets the maximum deformation of the grid (the
cell size can only become $\sim 2B_*$ times smaller than the initial
value). The value of $B_*$ cannot be arbitrarily large because the
grid has a ``numerical tension'' that prevents excessive deformation.

In Figure~\ref{fig:sfr2} we show the fraction of baryons in stars,
$f_{star}$, as a function of redshift for values of $B_*=8, 10, 12,
16$. It appears that the results are not sensitive to the value of
$B_*$, at least in the explored range. The only noticeable effect is a
slight delay in the formation of the first objects as $B_*$ increases.

\def\capfigi{Fraction of baryons in stars versus redshift for
  simulations with different values of the spatial resolution
  parameter, $B_*$. The two thick lines have $N_{box}=128$,
  $L_{box}=1$ \Mpc ($B_*=10, 16$). The thin solid and dotted lines have
  $N_{box}=64$, $L_{box}=1$ \Mpc ($B_*=10, 16$). The other thin lines
  have $N_{box}=64$, $L_{box}=0.5$ \Mpc ($B_*=8,10,12,16$). The spatial
  resolution is about $L_{box}/(N_{box}B_*)$.}
\placefig{
\begin{figure*}[thp]
\plotone{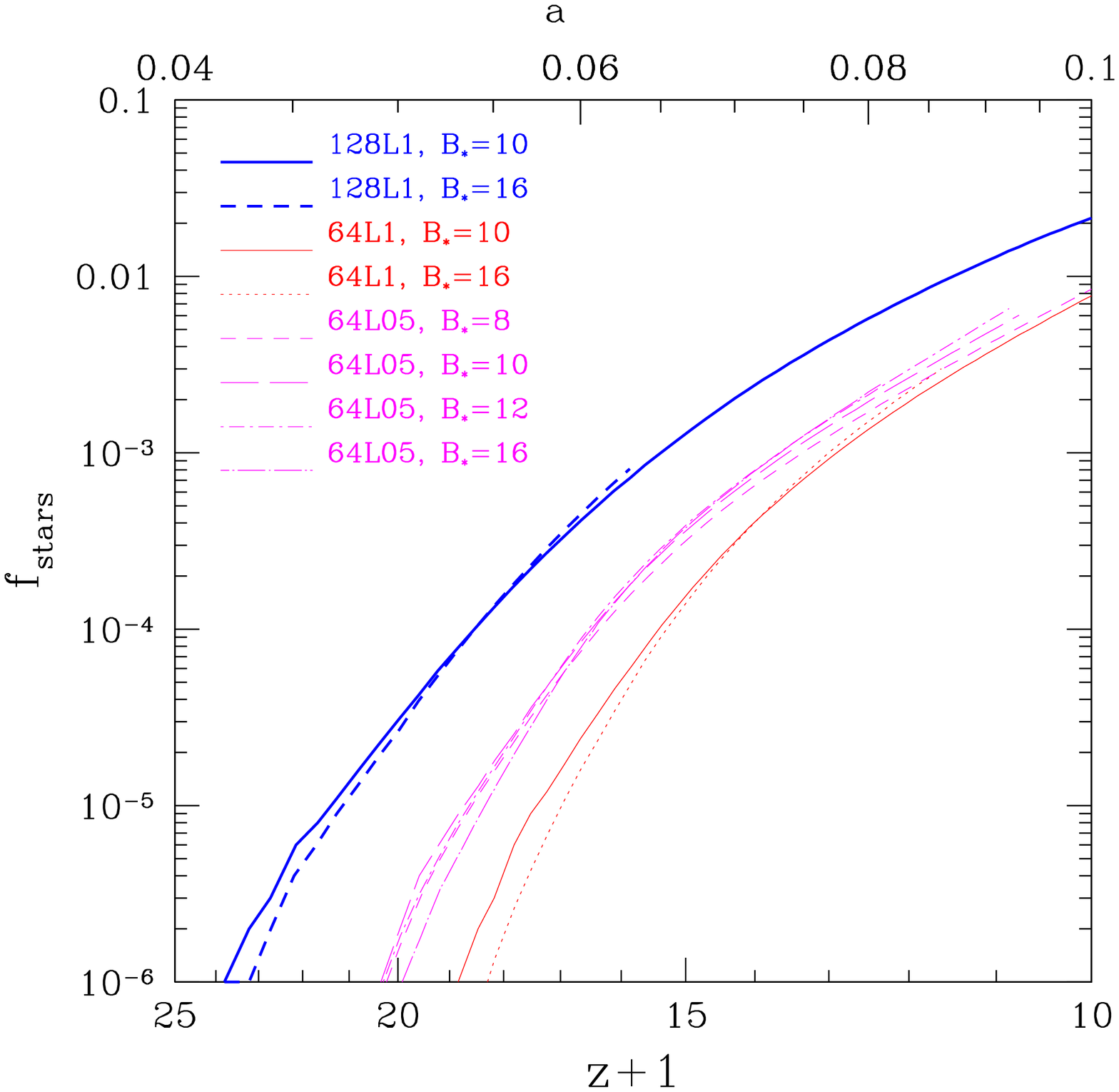}
\epsscale{1.}
\caption{\label{fig:sfr2}\capfigi}
\end{figure*}
}

\section{Discussion and Summary \label{sec:sumI}}

We have introduced a 3D cosmological code with radiative transfer
suited to simulate the formation of the first objects
self-consistently.  We use the ``Softened Lagrangian Hydrodynamics''
(SLH-P$^3$M) code \citep{Gnedin:95, Gnedin:96a, GnedinB:96,
  GnedinA:01} to solve the system of time-dependent equations of
motion for dark matter particles (P$^3$M algorithm), gas particles
(quasi-Lagrangian deformable mesh using the SLH algorithm), and
``stellar-particles'' formed using the Schmidt law in resolution
elements that sink below the numerical resolution of the code.  The
highest mass resolution that we achieve is $M_{DM}=4.93\times 10^3$
\Ms, and the highest spatial resolution is $156$ \pc comoving. We
solve the continuum radiative transfer with the OTVET approximation
\citep{GnedinA:01}, and we solve exactly the line radiative transfer
in the \H2 Lyman-Werner bands of the background radiation. Detailed
atomic and molecular physics is included, although we have not
included ``supernova feedback'' (work in progress).

Including ionization from secondary electrons in the simulation does
not produce significant changes of the IGM temperature or ionization
of H and He.  Secondary electrons are important when the ionizing
spectrum is hard and UV photons do not dominate the ionization rate.
X-rays (keV energies) in the background radiation are produced either
by Wolf-Rayet stars in \popII or by the much hotter \pop3 stars.
Nevertheless, the ionizing radiation background is dominated by the UV
photons in the spectral features produced by the redshifted resonant
\GI and \GII Ly$\alpha$ lines. In Figure~\ref{fig:rad1} we show three
examples of background (\ie, spatially averaged) radiation spectra:
for a \pop3 SED and \fesc$=1$, for a \pop3 SED and \fesc$=10^{-2}$,
and for a \popII SED and \fesc$=10^{-2}$. The spectrum is rather
complex, but we can identify spectral features associated with the
redshifted resonant \HI, \GI, and \GII Ly$\alpha$ lines, and the jumps
at the \HI, \GI, and \GII ionization thresholds. In the spectral
region between 11.2 - 13.6 eV, responsible for the photodissociation
of \H2, we solve the line radiative transfer using a grid with
$20,000$ frequency bins. An enlarged figure of this spectral region
would show more clearly the fine absorption features associated with
\H2 and resonant \HI line opacity in the IGM. For the rest of the
spectrum radiative transfer is solved for both the background and the
spatially varying radiation dividing the spectrum in 4 frequency
intervals [(i) optically thin between 1 - 11.2 eV, (ii) \HI ionizing
radiation, (iii) \GI ionizing radiation and (iv) \GII ionizing
radiation] resolved with 300 logarithmic bins altogether.

\def\capfigl{Spectrum of the background radiation, $J_\nu$. We show
  three representative examples: (i) using a \pop3 SED with \fesc$=1$
  (solid line); (ii) using a \pop3 SED with \fesc$=10^{-2}$ (dashed
  line); (iii) using a \popII SED with \fesc$=10^{-2}$ (dotted line).
  The dotted vertical line shows the \GI ionization frequency, and the dashed
  vertical line shows the \GII ionization frequency.}
\placefig{
\begin{figure*}[thp]
\plotone{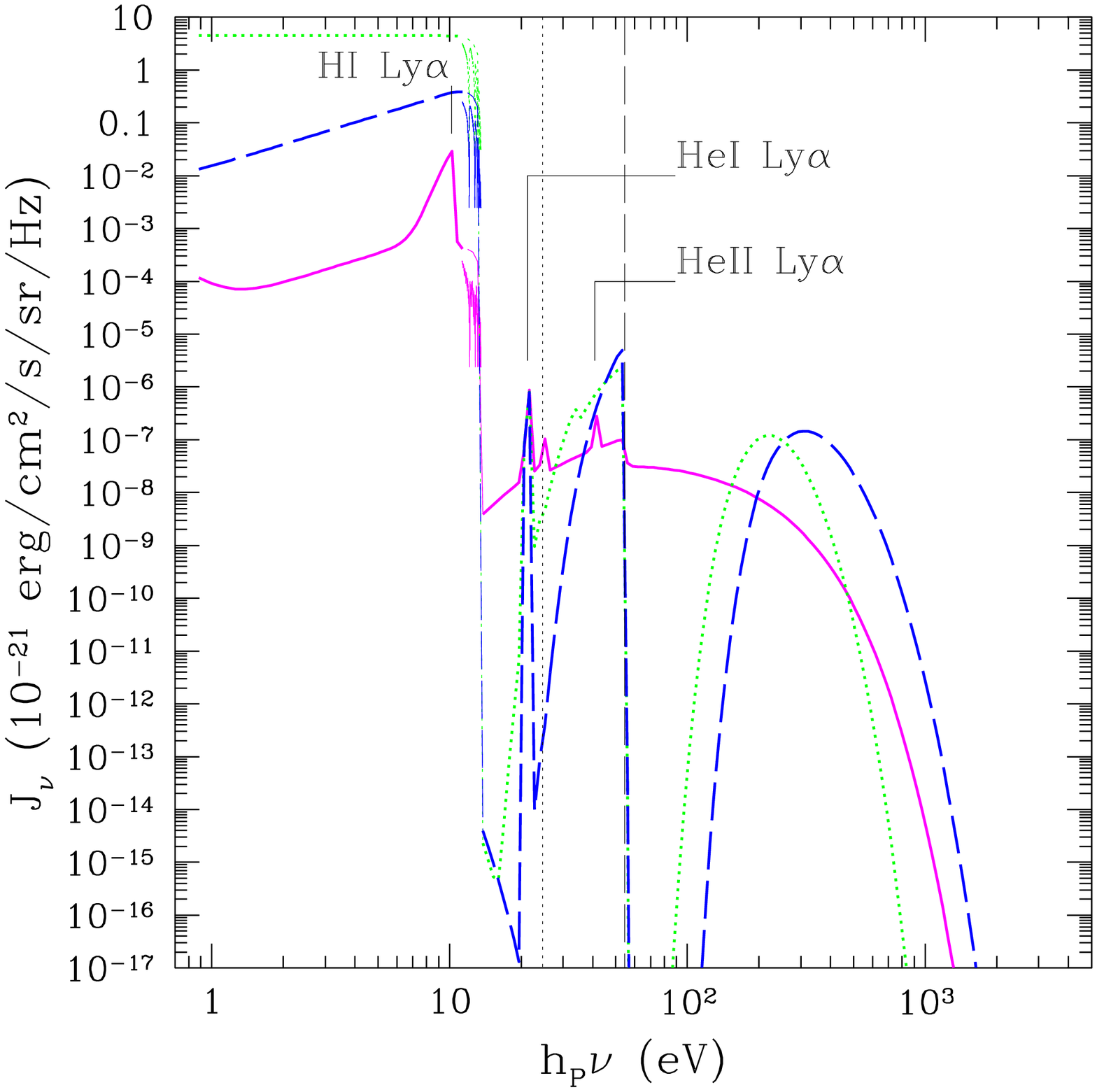}
\epsscale{1.}
\caption{\label{fig:rad1}\capfigl}
\end{figure*}
}

In Figure~\ref{fig:rendI} we show an example of 3D rendering of \HII
regions at $z=27.6$ from one of our simulations ($N_{box}=128$,
$L_{box}=1$ \Mpc) with radiative transfer. The small \HII regions
produced by \p3 objects (the first \pII objects form much later,
cooling by \HI Ly$\alpha$, at $z \simlt 18$) are clustered along the
dense filaments of the cosmic large scale structure.  In
Figure~\ref{fig:sfrI} we show the global SFR as a function of redshift
in the 256L1p3 run (thick solid line) compared to the same simulation
without radiative feedback effects (thin solid line) and to a case
where we do not allow the formation of any \p3 objects (thin dashed
line). It is evident that the bursting mode of SF appears in the
simulation with radiative feedback. The formation of \p3 objects is
not dramatically suppressed by radiative feedback effects.  In
Paper~II we study the nature of the feedback mechanism and the
dependence of the SFR on the free parameters (\fesc, $\epsilon_*$,
$\epsilon_{UV}$, and $g_\nu$) related to the unknown sub-grid physics
(stellar IMF, star formation efficiency, and ISM properties at
high-redshift).

\def\capfigm{Example of 3D rendering of \HII regions at $z=27.6$. The
  colors code the logarithm of \HII abundance. \p3 objects produce
  small \HII regions clustered along the large scale structure filaments.}
\placefig{
\begin{figure*}[thp]
\plotone{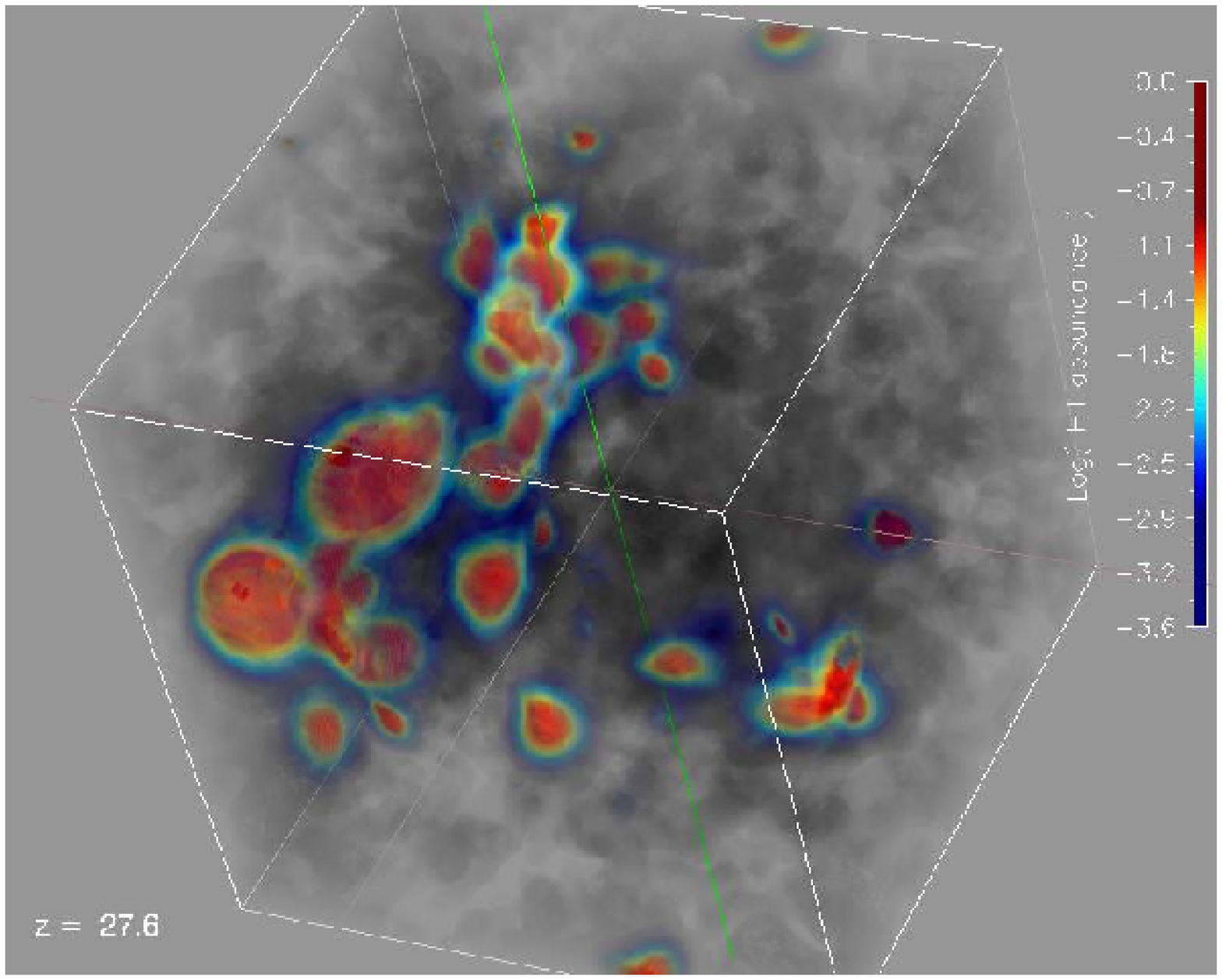}
\epsscale{1.}
\caption{\label{fig:rendI}\capfigm}
\end{figure*}
}

\def\capfign{Global SFR versus redshift. The thick solid line shows
  the 256L1p3 run ($N_{box}=256$, $L_{box}=1$ \Mpc, \pop3 SED, \fesc$=0.1$,
  $\epsilon_*=0.1$, and $(\epsilon_{UV}/4\pi)=2.5 \times 10^{-5}$) that
  includes continuum and line radiative transfer and all the
  physics discussed in this paper. As a comparison, we show the same
  simulation without radiative transfer (thin solid line)
  and a simulation ($N_{box}=64$, $L_{box}=1$ \Mpc) that forms only
  \pII objects since we did not include \H2 cooling and radiative
  transfer (thin dashed line).}
\placefig{
\begin{figure*}[thp]
\plotone{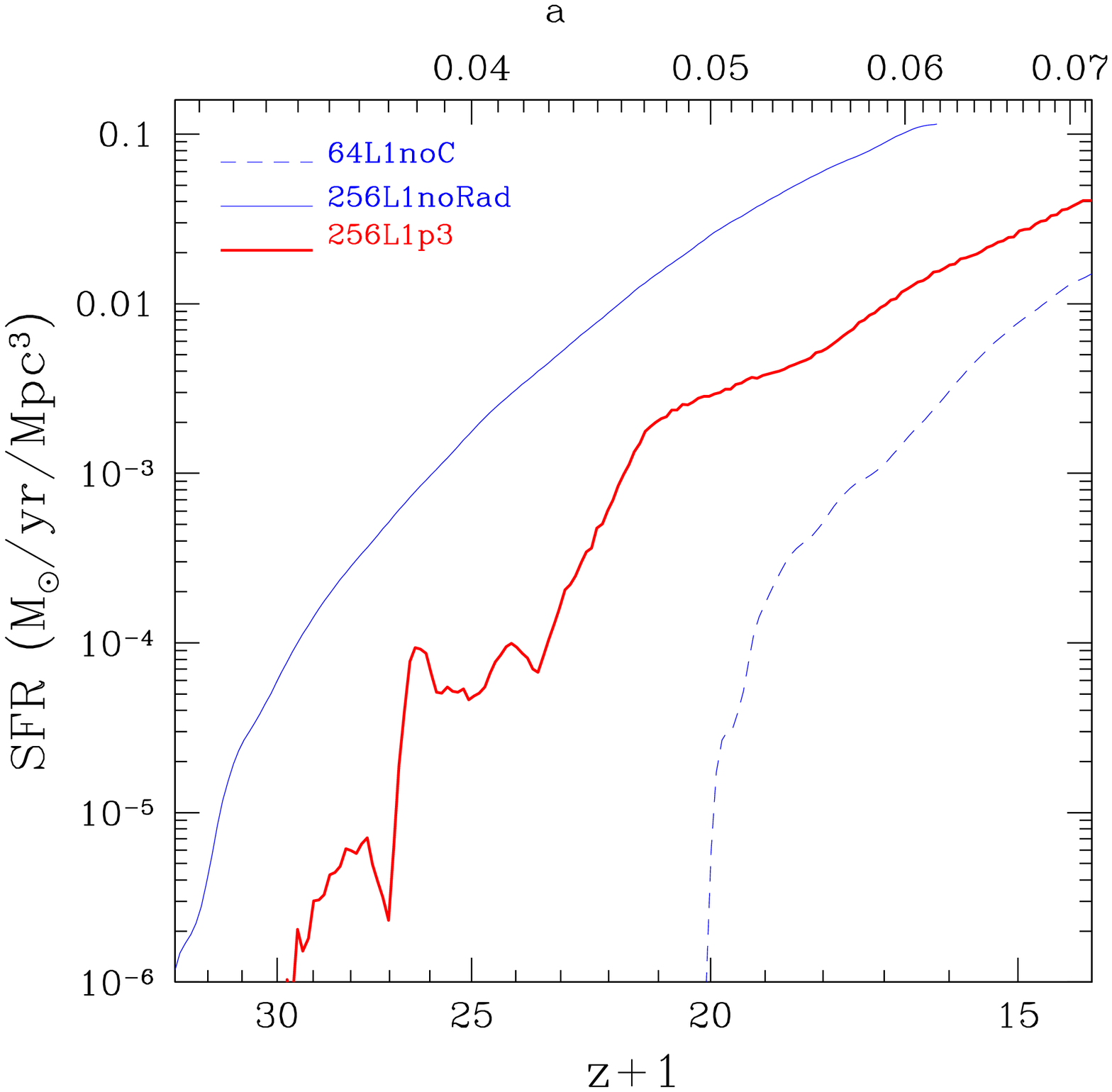}
\epsscale{1.}
\caption{\label{fig:sfrI}\capfign}
\end{figure*}
}

Two important processes are not yet included in the simulations: H$_2$
self-shielding and feedback from SN explosions.  In some of our
simulations with radiative transfer, we find that \H2 in the filaments
has a column density $N_{H_2} \ge 10^{15}$ cm\mm, sufficient to shield the
outgoing radiation emitted by embedded sources and the ingoing
radiation emitted by external sources.  We could crudely include the
effect of \H2 self-shielding for the outgoing radiation emitted by
each source embedded in the filaments, modifying the source spectrum
in the Lyman-Werner bands according to the 1D radiative transfer
calculations shown in Figure~5 in \cite{RicottiGS:01}. A complete
treatment of line radiative transfer would necessarily require some
approximation, since it is impossible to achieve the frequency
resolution needed for the exact calculation. We have decided not to
implement any of these two possibilities, based on the first results
of the simulations.  In Paper~II we will show that the SF is not
suppressed by the dissociating radiation, even if we neglect \H2
self-shielding. The inclusion of self-shielding will not affect the
global SFR in a significant way, since the primary feedback process that
regulates the star formation is not \H2 photodissociation.

Supernova explosions could be important sources of feedback, both of
heat and ionization. They could also produce a self-regulating global
star formation and contribute to the transport of metals in the low
density IGM.  Unfortunately, we do not yet understand the correct
modeling of their dynamical and thermal effects on the ISM/IGM. An
approximate treatment of SN mechanical and thermal energy input is
currently implemented in the code. We did not use this feature of the
code, because we prefer to study the effects of radiative feedback and
mechanical feedback from SN explosions separately. Moreover, we are
not confident that the treatment of SN explosions in the code is
appropriate, given the complexity of ISM in galaxies. The
investigation of the effects of mechanical feedback from SNe
explosions will be the subject of our future work.

\acknowledgements 

This work was supported by the Theoretical Astrophysics program at the
University of Colorado (NASA grant NAG5-7262). The simulations
presented in this paper were performed using SGI/CRAY Origin 2000
array at the National Center for Supercomputing Applications (NCSA).

\appendix
\section{Secondary Ionization and Heating from X-rays \label{ap:sec}}

For photoelectron energies $E_0^i \gg 100$ eV we use the fits given by
\cite{ShullVan:85}.  For lower energy photoelectrons we have made fits
to the Monte Carlo results shown in Figure~3 of the aforementioned
paper. The functional form of the fits allow us to integrate in photon
frequency, factoring out the dependency on $x_e$. We pre-calculate the
integrals and store the results in three-dimensional tables
for each frequency bin in order to avoid computationally intensive
calculations.

The fraction of secondary ionizations per primary electron of energy
$E_0^i$ is $\Phi^j(E_0^i,x_e)$, where $E_0^i =
h_p \nu - I^i$ with $I^i$ ground-state ionization potential of the
species $i$ and  
\begin{equation}
\Phi^j (E_0^i,x_e)= \left(E_0^i \over I^j\right)[y_1(j)f_1(E_0^i) - y_2(j)f_2(E_0^i)],
\label{eq:ap1}
\end{equation}
where
\begin{eqnarray}
f_1 =
\begin{cases}
0 & \text{if $E_0^i < 28$ eV}\\
1 & \text{if $E_0^i \ge 28$ eV}\\
\end{cases}
~~~f_2 = 
\begin{cases}
0 & \text{if $E_0^i < 28$ eV}\\
(28~{\rm eV} /E_0^i)^{0.4} & \text{if $E_0^i \ge 28$ eV}.\\
\end{cases}
\end{eqnarray}
Here, since we neglect ionization and excitation of \GII from
energetic primary electrons, $i = \HI, \GI, \GII$ and $j = \HI, \GI$.
Since a fraction of the energy of primary photoelectrons goes to
secondary ionizations, the heating rates from \HI and
\GI photoionizations are less efficient by the complementary factor,
\begin{equation}
E_h (E_0^i,x_e)= E_0^i [1- y_1(heat)f_1(heat) + y_2(heat)f_2(heat)],
\label{eq:ap2}
\end{equation}
where
\begin{eqnarray}
f_1 =
\begin{cases}
0 & \text{if $E_0^i < 11$ eV}\\
1 & \text{if $E_0^i \ge 11$ eV}\\
\end{cases}
~~~f_2 = 
\begin{cases}
0 & \text{if $E_0^i < 11$ eV}\\
(11~{\rm eV} /E_0^i)^{0.7} & \text{if $E_0^i \ge 11$ eV}.\\
\end{cases}
\end{eqnarray}
The high energy limits of equations~(\ref{eq:ap1})--(\ref{eq:ap2}) are the
functions $y_1$ that are given by \cite{ShullVan:85}, and $y_2$ has the
form,
\begin{eqnarray}
y_1 &=& C(1-x_e^b)^c\\
y_2 &=& Cx^a(1-x_e^b)^c ,
\label{eq:y2}
\end{eqnarray}
with $0 \le x_e \le 1$. The coefficients for equation~(\ref{eq:y2})
are listed in Table~\ref{tab:A} and the fits are shown in Figure~\ref{fig:A}.

\def\tabtwo{
\begin{deluxetable}{lcccc}

\tablecaption{Fitting Coefficients.\label{tab:A}}
\tablewidth{0pt}
\tablehead{
\colhead{Parameter} & \colhead{$C$} & \colhead{$a$} & \colhead{$b$} & \colhead{$c$}
 }
 \startdata
$y_1(\HI)$  & 0.3908 & -- & 0.4092 & 1.7592 \\
$y_1(\GI)$  & 0.0554 & -- & 0.4614 & 1.6660 \\
$y_1(heat)$ &  1.0   & -- & 0.2663 & 1.3163 \\

$y_2(\HI)$  & 0.6941 & 0.2 & 0.38 & 2.0 \\
$y_2(\GI)$  & 0.0984 & 0.2 & 0.38 & 2.0 \\
$y_2(heat)$ & 3.9811 & 0.4 & 0.34 & 2.0 \\
\enddata
\end{deluxetable}
}
\placefig{\tabtwo} 

\def\capfigo{(Left) Fraction of energy deposited
  as \HI ionization per primary electron of energy $E_0^i$, as a
  function of the electron fraction $x_e$. The data points and curves
  are for $E_0^i=28, 50, 100, 200$ eV from bottom to top. (Right) Fraction of
  energy deposited as heat per primary electron of energy $E_0^i$, as
  a function of $x_e$. The data points and curves are for $E_0^i=2,
  11, 28, 50, 200$ eV from top to bottom. }  
\placefig{
\begin{figure*}[thp]
\plottwo{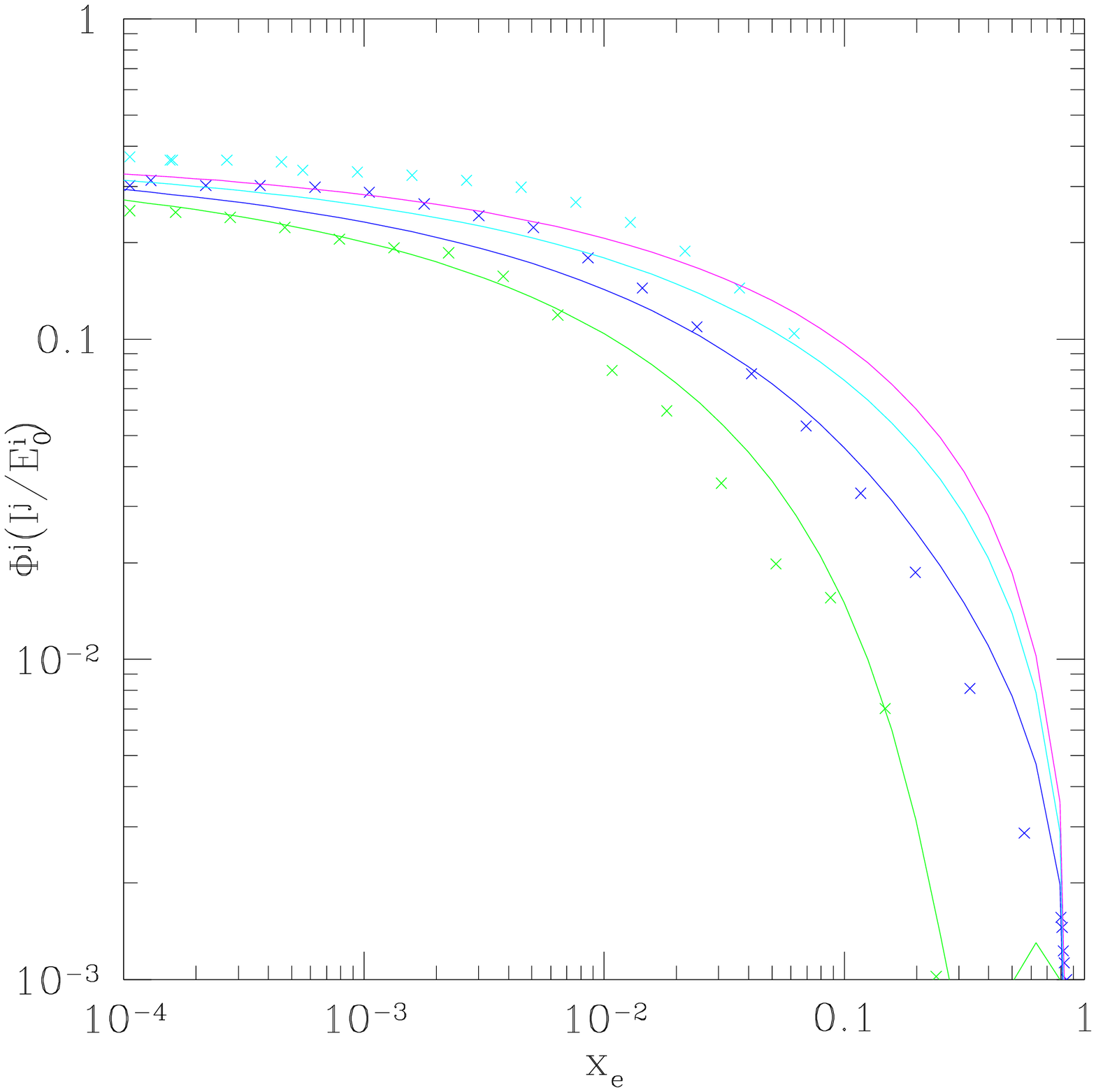}{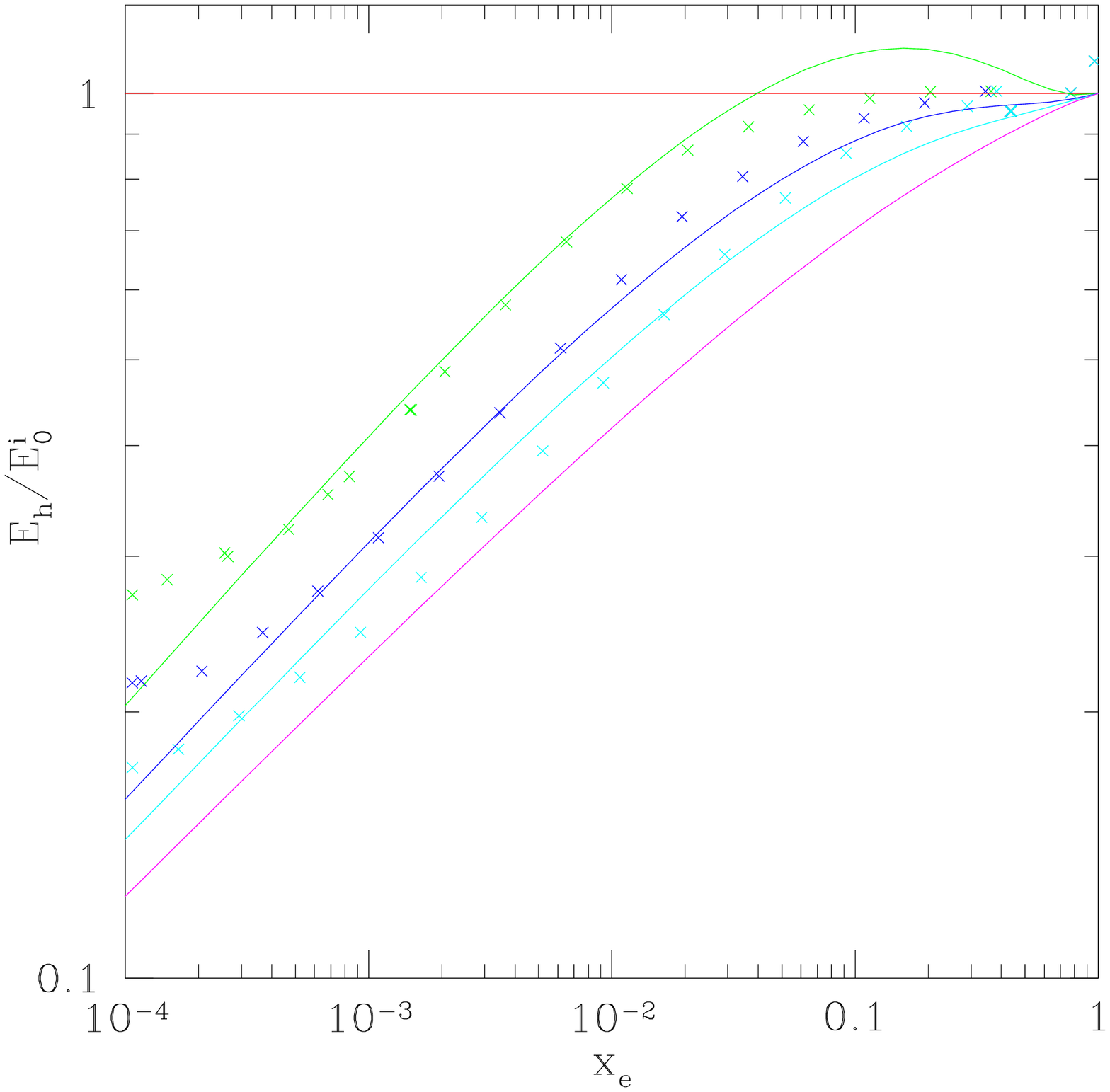}
\caption{\label{fig:A}\capfigo}
\end{figure*}
}

\section{Group-finding Algorithm\label{app:B}}

The DENMAX algorithm \citep{Bertschinger:91} identifies halos as
maxima of the smoothed density field. The algorithm assigns each
particle in the simulation to a group (halo) by moving each particle
along the gradient of the density field until it reaches a local
maximum. Unbound particles are then removed from the group. 

The results of DENMAX depend on the degree of smoothing used to define
the density field. A finer resolution in the density field will split
large groups into smaller subunits and vice versa. This arbitrariness in
results is a common problem of any group-finding algorithm, and it is
not only a numerical problem but often a real physical ambiguity.
Especially at high redshift, since the merger rate is high, it is
difficult to identify or define a single galaxy halo.
  
Here we compare the mass function obtained using DENMAX with different
smoothing scales $L_G = L_{box}/G$ with the Press-Schechter formalism.
The main aim is to choose the parameter $G$ consistently in order to
compare the results from simulations with an arbitrary number of cells
in the box $N_{box}^3$. The mass of a DM particle is $M_p = 8.26
\times 10^{10} [(L_{box}/1 h^{-1}{\rm
  Mpc})^3/N_{box}^3(\Omega_m/0.3)]$ \Ms and the comoving space
resolution $L_{res}=L_{box}/(B_*N_{box})$.

If we impose that each group has a minimum of $N_p$ DM particles, we have $L_G
\ge N_p^{1/3}L_{res}$, which implies,
\begin{equation}
G \le {B_* N_{box} \over N_{p}^{1/3}} < {B_* N_{box}},
\label{eq:B1}
\end{equation}
where $B_*$ is the resolution parameter. The inequality on the
right-hand side is obtained by imposing $N_p \ge 1$.  Another constraint
on $G$ can be obtained using a smoothing length comparable to the
halo virial radius $L_G \approx R_{vir}\approx 3.4~{\rm kpc} (M/10^6
{\rm M}_\odot)^{1/3}$ of the smallest object that we want to find:
\begin{equation}
G \approx {6.75 N_{box} \over N_{p}^{1/3}}.
\label{eq:B2}
\end{equation}
Note that the mass of the halo is $M = N_p M_{res}$.
Combining equations~(\ref{eq:B1}) and (\ref{eq:B2}), we get
\begin{equation}
{G N_{p}^{1/3} \over N_{box}} \approx 6.75 \le B_*.
\label{eq:B3}
\end{equation}

In Figure~\ref{fig:B} we show the mass function compared to the
Press-Schechter formalism with a top-hat window function.  Better
agreement with the Press-Schechter formula is obtained using $G
\approx (4-5) N_{box}$ [\ie, using $N_p \approx (2-5)$ in
equation~(\ref{eq:B3})].

\def\capfigp{Comparison of the mass functions (histograms) varying
  the smoothing parameter $G$ and the Press-Schechter formalism with a
  top-hat filter (thick solid line).  (Top-left) 64L05 and 64L1 at
  $z=11$ with $G=320, 400$.  (Top-right) 64L05 at $z=19$ with
  $G=80,160,320$.  (Bottom-left) 128L05 and 128L1 at $z=12$ with
  $G=320,400$.  (Bottom-right) 128L1 with $G=400$ and 256L1 with
  $G=1000$ at $z=15.13$.}
\placefig{
\begin{figure*}[thp]
\epsscale{1.}
\plottwo{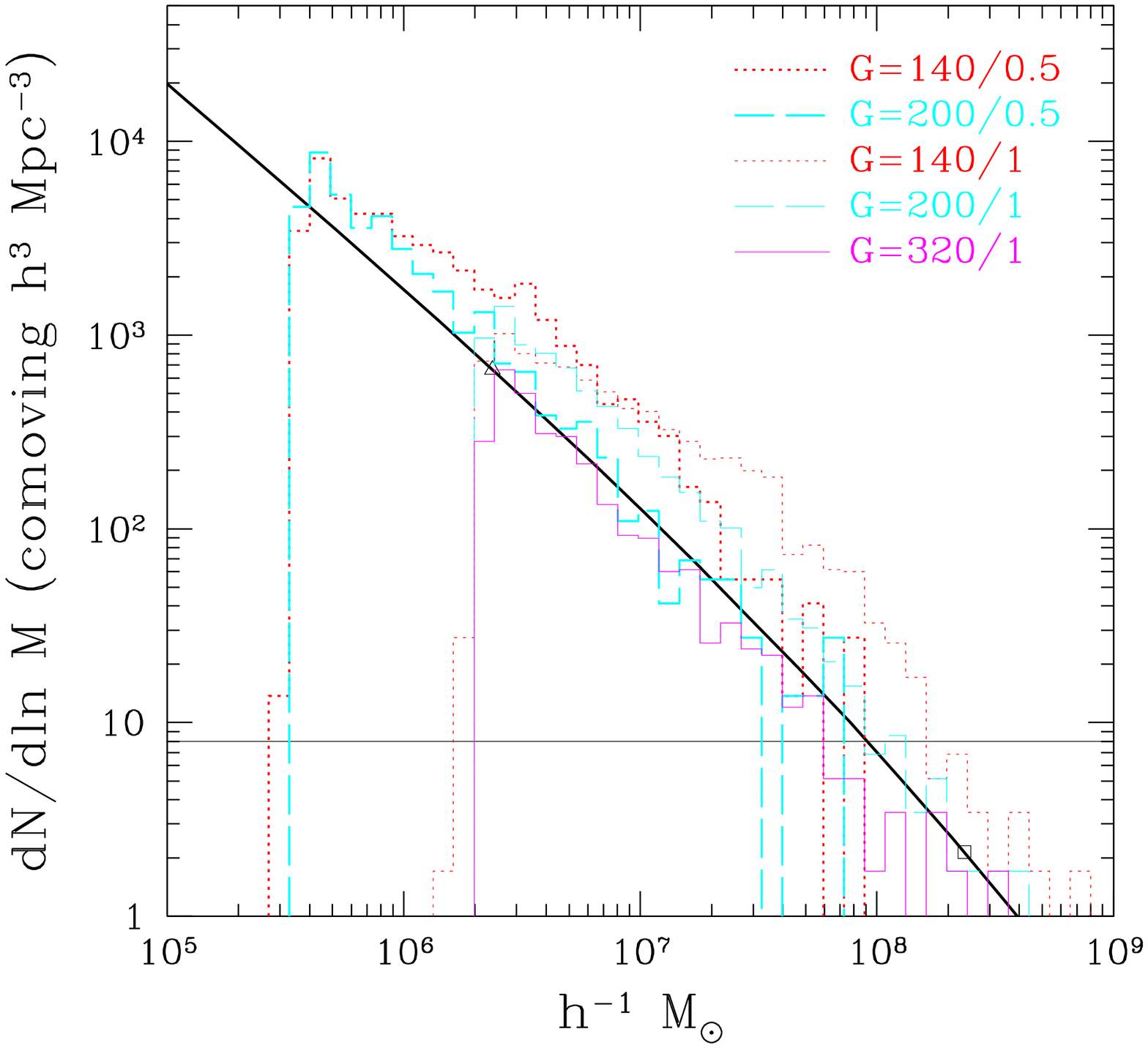}{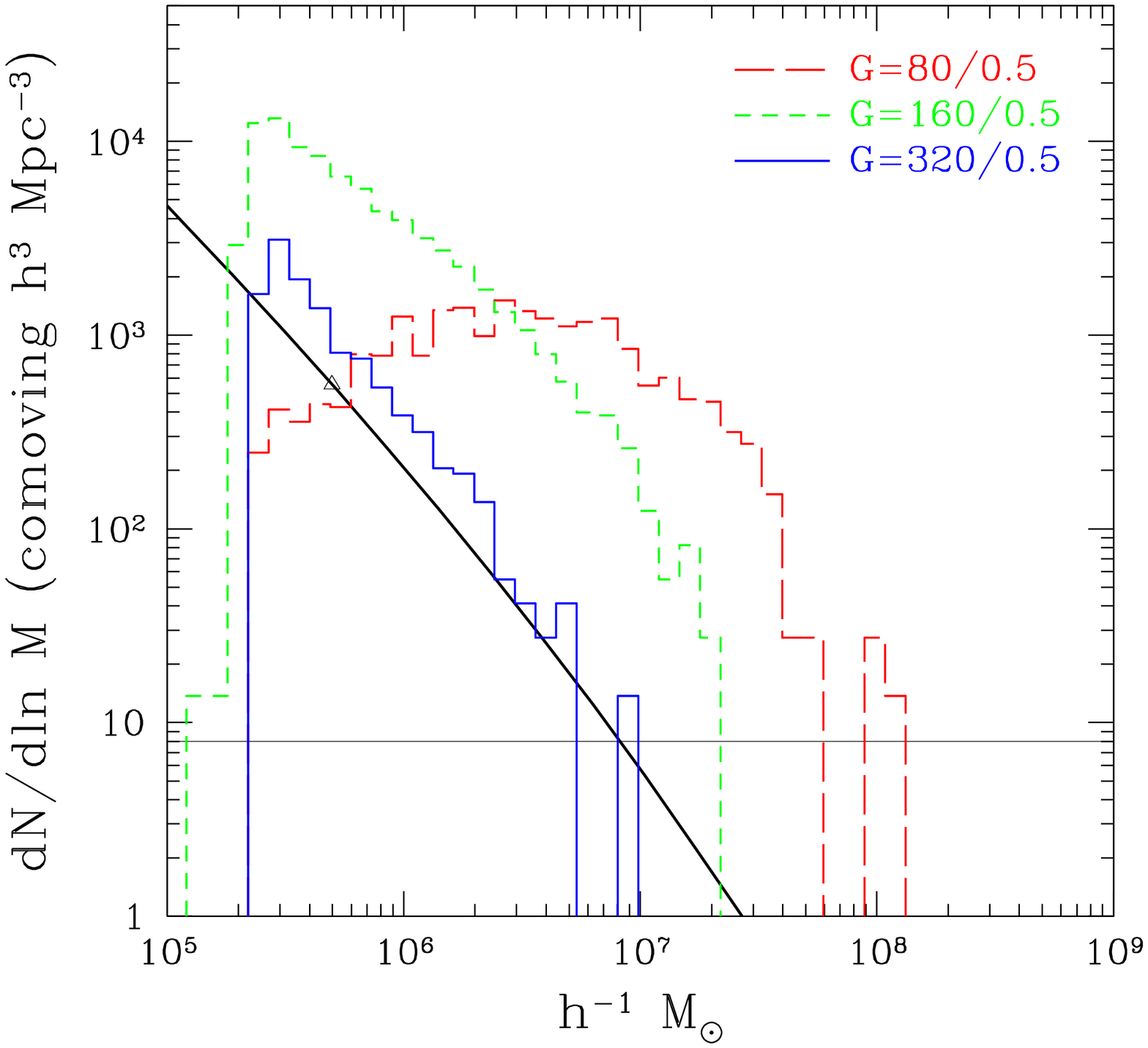}
\epsscale{2.2}
\plottwo{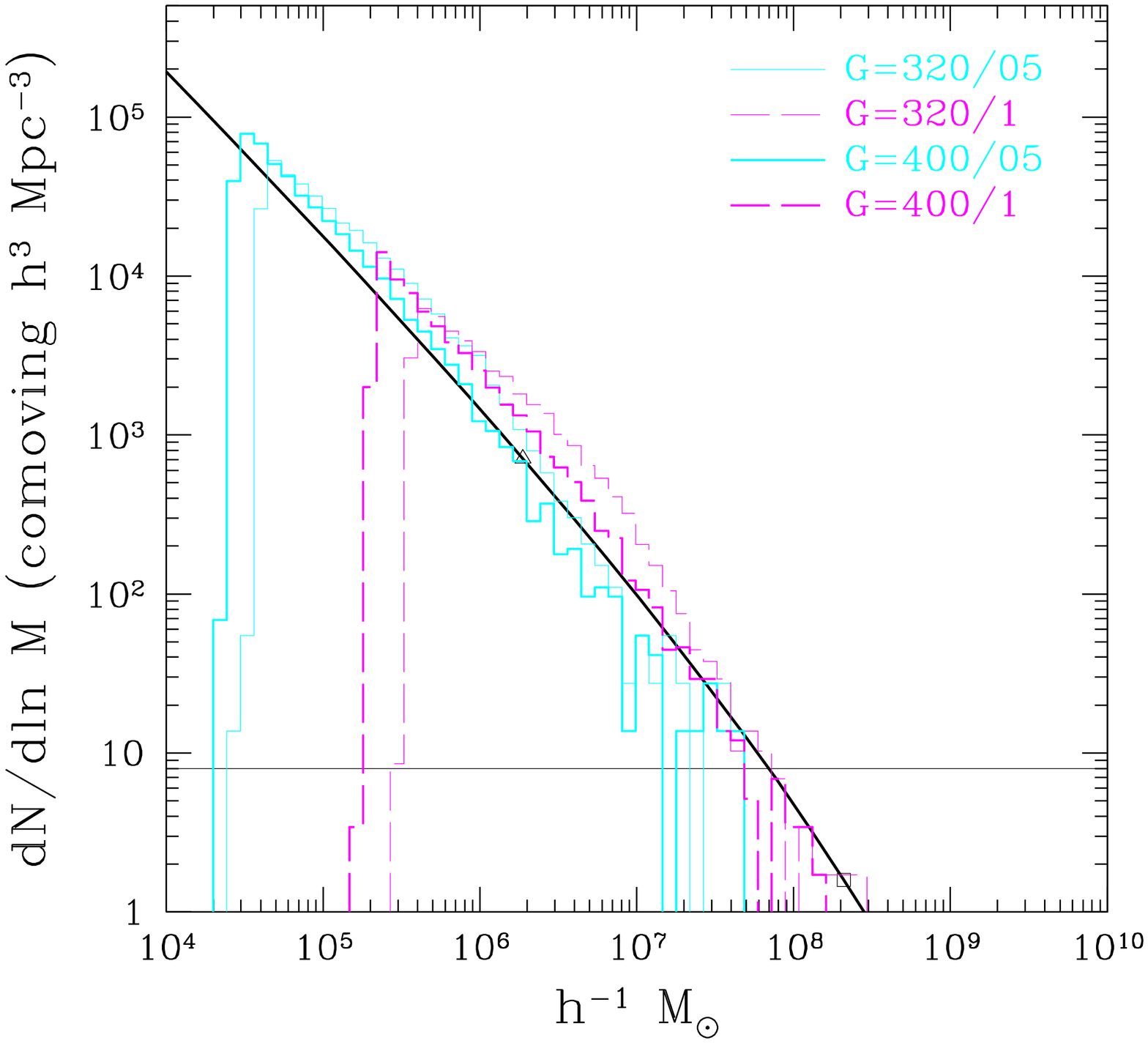}{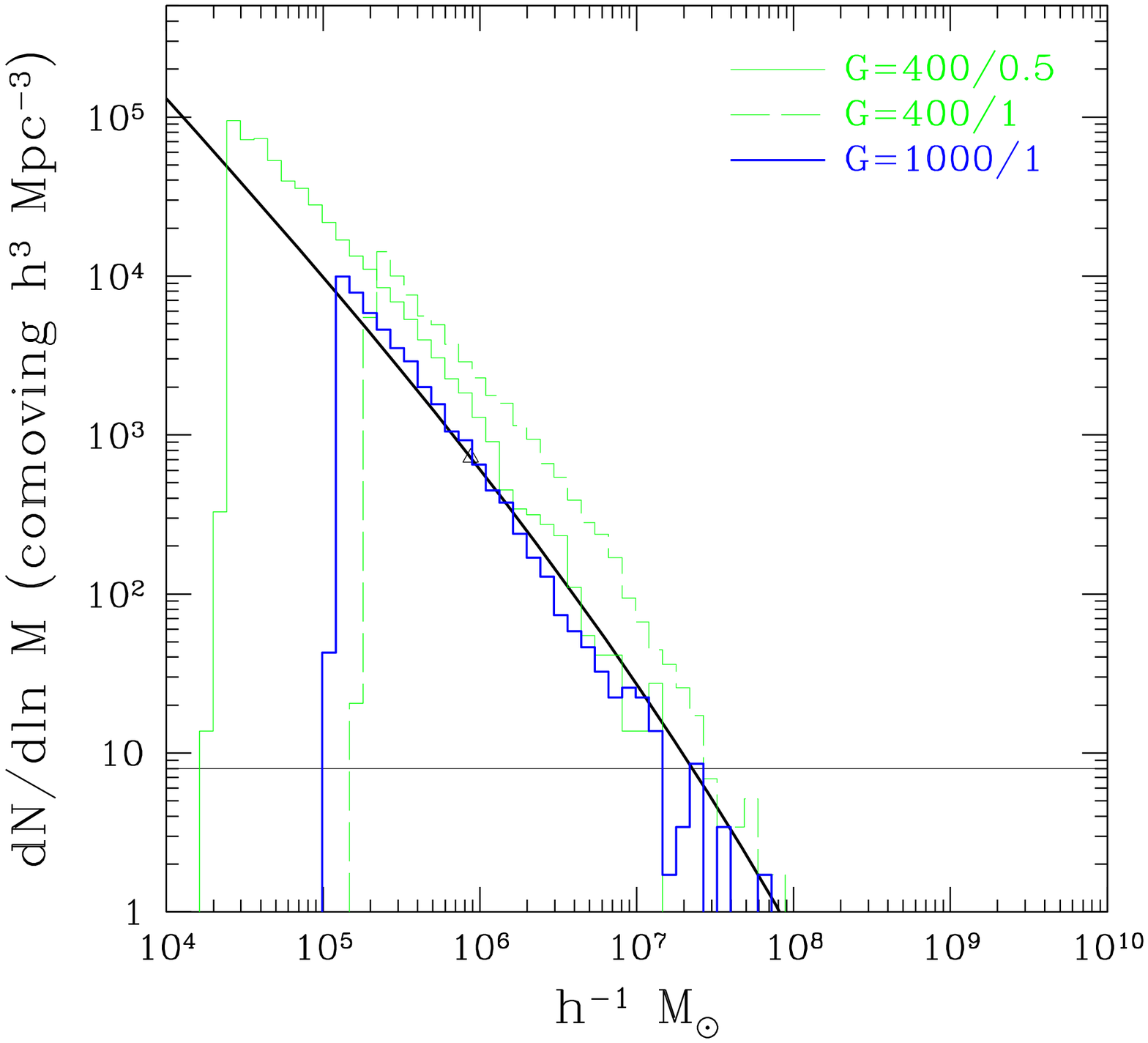}
\caption{\label{fig:B}\capfigp}
\end{figure*}
}
\clearpage
\bibliographystyle{/home/origins/ricotti/Latex/TeX/apj}
\bibliography{/home/origins/ricotti/Latex/TeX/archive}

\vskip 2truecm

\placefig{\end{document}}

\clearpage

\newcounter{figurecap}
\setcounter{figurecap}{0}

\begin{center}
\bf Figure Captions
\end{center}

\refstepcounter{figurecap}
Fig.\ \thefigurecap---\label{fig:ps}\capfiga

\refstepcounter{figurecap}
Fig.\ \thefigurecap---\label{fig:plot}\capfigb

\refstepcounter{figurecap}
Fig.\ \thefigurecap---\label{fig:rates}\capfigc

\refstepcounter{figurecap}
Fig.\ \thefigurecap---\label{fig:pop3}\capfigd

\refstepcounter{figurecap}
Fig.\ \thefigurecap---\label{fig:res}\capfige

\refstepcounter{figurecap}
Fig.\ \thefigurecap---\label{fig:sfr1}\capfigf

\refstepcounter{figurecap}
Fig.\ \thefigurecap---\label{fig:colMF}\capfigg

\refstepcounter{figurecap}
Fig.\ \thefigurecap---\label{fig:sf}\capfigh

\refstepcounter{figurecap}
Fig.\ \thefigurecap---\label{fig:sfr2}\capfigi

\refstepcounter{figurecap}
Fig.\ \thefigurecap---\label{fig:rad1}\capfigl

\refstepcounter{figurecap}
Fig.\ \thefigurecap---\label{fig:rendI}\capfigm

\refstepcounter{figurecap}
Fig.\ \thefigurecap---\label{fig:sfrI}\capfign

\refstepcounter{figurecap}
Fig.\ \thefigurecap---\label{fig:A}\capfigo

\refstepcounter{figurecap}
Fig.\ \thefigurecap---\label{fig:B}\capfigp

\clearpage

\tabone

\tabtwo

\end{document}